\documentclass[preprint,floatfix]{revtex4}

\pdfoutput=1

\usepackage{pdfpages}
\usepackage{graphicx} 
\usepackage{dcolumn} 
\usepackage{amsmath}
\usepackage{amssymb}
\usepackage{bm}
\usepackage{times}
\usepackage{ulem}
\usepackage[justification=raggedright]{caption}

  \usepackage{boondox-cal}

\usepackage{subfigure}
\usepackage{tabularx}
\usepackage{appendix}

\usepackage{amstext}
\usepackage{array}

\usepackage{bbm}
\usepackage{blkarray}

\definecolor{wine-stain}{rgb}{0.5,0,0} 
\definecolor{bblue}{rgb}{0,0.0,0.5} 

\usepackage[colorlinks=true
,urlcolor=blue
,anchorcolor=blue
,citecolor=blue 
,filecolor=blue
,linkcolor=wine-stain
,menucolor=blue
,pagecolor=blue
,linktocpage=true
,pdfproducer=medialab
,linktoc=all 
]{hyperref}

\usepackage{booktabs}  

\allowdisplaybreaks

\newcommand{\ncmd}{\newcommand}

\ncmd{\lt}{\left}
\ncmd{\rt}{\right}
\ncmd{\tr}[1]{~\mbox{tr}\lt\{ {#1}\rt\}}
\ncmd{\half}{\frac{1}{2}}
\ncmd{\eps}{\epsilon}
\ncmd{\veps}{\varepsilon}
\ncmd{\dgr}{\dagger}
\ncmd{\sig}{\bar \sigma}
\ncmd{\gam}{\gamma}
\ncmd{\rtarw}{\rightarrow}
\ncmd{\Rt}{\Rightarrow}
\ncmd{\abs}[1]{\lt\cb{#1}\rt\cb}
\ncmd{\avg}[1]{\lt\lb{#1}\rt\rb}
\ncmd{\sgn}[1]{\mbox{sgn}\lt(#1\rt)}
\ncmd{\kap}{\kappa}
\ncmd{\wtil}[1]{\widetilde{#1}}
\ncmd{\thrfr}{\therefore}
\ncmd{\eq}[1]{Eq. \eqref{#1}}
\ncmd{\fig}[1]{Fig. \ref{#1}}
\ncmd{\ordr}[1]{\mathcal{O}\lt(#1\rt)}
\ncmd{\dsty}{\displaystyle}
\ncmd{\alert}[1]{\color{red}{#1}}
\ncmd{\mc}{\mathcal}
\ncmd{\mbf}[1]{\mathbf{#1}}
\ncmd{\Deriv}[2]{\frac{d{#1}}{d{#2}}}
\ncmd{\ParDeriv}[2]{\frac{\partial{#1}}{\partial{#2}}}
\ncmd{\step}[1]{\Theta\lt(#1\rt)}
\ncmd{\td}{\tilde} 
\ncmd{\what}{\widehat}

\ncmd{\hphi}{\hat \phi} 
\ncmd{\hpi}{\hat \pi} 
\ncmd{\hK}{\hat K} 
\ncmd{\hL}{\hat L} 
\ncmd{\hU}{\hat U} 
\ncmd{\hQ}{\hat Q} 

\ncmd{\sL}{ \sqrt{L} } 

\ncmd{\cs}{{\cal s}}
\ncmd{\qq}{(q^T q)}
\ncmd{\cq}{{\cal q}}
\ncmd{\bH}{{\bf H}}
\ncmd{\bpp}{\kappa}
\ncmd{\app}{(1-\kappa)}
\ncmd{\bG}{{\bf G}}
\ncmd{\bC}{{\bf C}}

\ncmd{\bqa}{\begin{eqnarray}} 
\ncmd{\eqa}{\end{eqnarray}}

\ncmd{\nn}{\nonumber \\}
\ncmd{\comment}[1]{{\color{red}{#1}}}
\definecolor{new_color}{RGB}{50,155,0}

\ncmd{\vrho}{\varrho}

\ncmd{\ha}{\frac{1}{2}}

\ncmd{\lb}{\big<}
\ncmd{\rb}{\big>}
\ncmd{\cb}{\big|}
\ncmd{\cH}{{\cal H}}
\ncmd{\lc}{l_c}
\ncmd{\Lc}{l_c}
\ncmd{\ta}{\tilde \alpha}

\ncmd{\bt}{\bar t}
\ncmd{\bp}{\bar p}

\ncmd{\rot}{\br_A, \br_B}
\ncmd{\rto}{\br_B, \br_A}

\ncmd{\PhiAi}{\Phi^A_{~~i}}
\ncmd{\PhiAj}{\Phi^A_{~~j}}
\ncmd{\PhiAk}{\Phi^A_{~~k}}
\ncmd{\PhiBi}{\Phi^B_{~~i}}
\ncmd{\PhiBj}{\Phi^B_{~~j}}
\ncmd{\PhiBk}{\Phi^B_{~~k}}

\ncmd{\PiiA}{\Pi^i_{~~A}}
\ncmd{\PijA}{\Pi^j_{~~A}}
\ncmd{\PikA}{\Pi^k_{~~A}}

\ncmd{\PiiB}{\Pi^i_{~~B}}
\ncmd{\PijB}{\Pi^j_{~~B}}
\ncmd{\PikB}{\Pi^k_{~~B}}

\ncmd{\hPhi}{\hat \Phi}
\ncmd{\hPi}{\hat \Pi}
\ncmd{\hG}{{\hat {\bf G}}}
\ncmd{\hH}{\hat H}
\ncmd{\hJ}{\hat J}

\ncmd{\GLL}{GL(L, \mathbb{R}) }
\ncmd{\SLL}{SL(L, \mathbb{R}) }

\ncmd{\bk}{{\bf k}}
\ncmd{\br}{{\bf r}}

\ncmd{\qw}{q^{(w)}} 
\ncmd{\qo}{q^{'}} 
\ncmd{\qt}{q^{(2)}}

\ncmd{\tds}{S}
\ncmd{\tdt}{T }
\ncmd{\tdp}{P}
\ncmd{\tdX}{X}
\ncmd{\tdY}{Y}

\ncmd{\hd}{\frac{r'}{2}}

\ncmd{\hPhiAi}{\hat \Phi^A_{~~i}}
\ncmd{\hPhiAj}{\hat \Phi^A_{~~j}}
\ncmd{\hPhiAk}{\hat \Phi^A_{~~k}}
\ncmd{\hPhiBi}{\hat \Phi^B_{~~i}}
\ncmd{\hPhiBj}{\hat \Phi^B_{~~j}}
\ncmd{\hPhiBk}{\hat \Phi^B_{~~k}}

\ncmd{\hPiiA}{\hat \Pi^i_{~~A}}
\ncmd{\hPijA}{\hat \Pi^j_{~~A}}
\ncmd{\hPikA}{\hat \Pi^k_{~~A}}
\ncmd{\hPiiB}{\hat \Pi^i_{~~B}}
\ncmd{\hPijB}{\hat \Pi^j_{~~B}}
\ncmd{\hPikB}{\hat \Pi^k_{~~B}}

\ncmd{\cA}{{\cal A}}
\ncmd{\cB}{{\cal B}}
\ncmd{\cC}{{\cal C}}
\ncmd{\cD}{{\cal D}}

\ncmd{\cG}{{\cal G}}
\ncmd{\cGI}{{\cal G}^{-1}}

\ncmd{\cGij}{{\cal G}^i_{~~j}}
\ncmd{\cGji}{{\cal G}^j_{~~iå}}

\ncmd{\cGIij}{({\cal G}^{-1})^i_{~~j}}
\ncmd{\cGIji}{({\cal G}^{-1})^j_{~~iå}}

\ncmd{\cstrdual}{\cb q, P_1, P_2 \rb}
\ncmd{\cstrdualc}{\cb q, P_1, P_2, X \rb}

\ncmd{\cstr}{\cb s, t_1, t_2 \rb}
\ncmd{\cstrp}{\cb s', t_1', t_2' \rb}

\ncmd{\czr}{ \cb 0 \rb } 
\ncmd{\lzc}{ \lb 0 \cb }

\ncmd{\cepr}{ \cb q, \phi, \varphi \rb^{'} }
\ncmd{\ketqspt}{\cb q,s,p_1,t_1,p_2,t_2 \rb}

\ncmd{\tx}{\td x}
\ncmd{\ty}{\td y}

\ncmd{\cphir}{\cb \phi \rb}
\ncmd{\lphic}{\lb \phi \cb}

\ncmd{\ctr}{\cb T \rb}
\ncmd{\ltc}{\lb T \cb}
\ncmd{\cpr}{\cb p \rb}
\ncmd{\lpc}{\lb p \cb}

\ncmd{\cTr}{\cb {\cal T} \rb}

\ncmd{\ccr}{\cb \chi \rb}
\ncmd{\lcc}{\lb \chi \cb}

\ncmd{\cS}{{\cal S}}
\ncmd{\tcS}{\tilde { \cal S}}

\ncmd{\sqg}{\sqrt{|g(x)|}}
\ncmd{\gmn}{E_{\mu i}}

\ncmd{\emi}{E_{\mu i}}
\ncmd{\emj}{E_{\mu j}}
\ncmd{\eni}{E_{\nu i}}
\ncmd{\enj}{E_{\nu j}}
\ncmd{\de}{|E(x)|}

\ncmd{\gemn}{g_{E,\mu \nu}}
\ncmd{\gemnp}{g_{E',\mu \nu}}

\ncmd{\bgmn}{\bar E_{\mu i}}
\ncmd{\Fs}{F(\sigma)}
\ncmd{\cJ}{ {\cal J}\left(E',\sigma';E,\sigma \right)  }
\ncmd{\Si}{ \Sigma \left(E',\sigma';E,\sigma \right)  }

\ncmd{\grs}{g_{\alpha \beta}}
\ncmd{\pmn}{\pi^{\mu i}}
\ncmd{\prs}{\pi^{\alpha \beta}}
\ncmd{\fs}{ \frac{3}{\sqrt{2}} \sigma }
\ncmd{\fst}{ 3 \sqrt{2} \sigma }

\ncmd{\ee}{entanglement entropy }

\ncmd{\cT}{{\cal T} }
\ncmd{\cP}{{\cal P} }
\ncmd{\cV}{{\mathbcal V} }
\ncmd{\cW}{{\cal W} }

\ncmd{\ccw}{\cos( 2 \omega \delta) }
\ncmd{\ssw}{\sin (2 \omega \delta) }
\ncmd{\bs}{{\bf s}}

\makeatletter
\newcommand*{\rom}[1]{\expandafter\@slowromancap\romannumeral #1@}
\makeatother

\begin{document}

\title{
Massless graviton in a model of quantum gravity with emergent spacetime
}

\author{Sung-Sik Lee\\
\vspace{0.3cm}
{\normalsize{Department of Physics $\&$ Astronomy, McMaster University, Hamilton ON, Canada}}
\vspace{0.2cm}\\
{\normalsize{Perimeter Institute for Theoretical Physics, Waterloo ON, Canada}}
}

\date{\today}

\begin{abstract}

In the model of quantum gravity proposed in 
JHEP {\bf 2020}, 70 (2020), 
dynamical  spacetime arises
as a collective phenomenon of underlying quantum matter.
Without a preferred decomposition of the Hilbert space,
the signature, topology and geometry of an emergent spacetime 
depend upon how the total Hilbert space is partitioned
into local Hilbert spaces.
In this paper, it is shown that 
the massless graviton 
emerges in the spacetime realized from a Hilbert space 
decomposition 
that supports a collection of largely unentangled local clocks.

\end{abstract}

\maketitle

\newpage

{
\hypersetup{linkcolor=bblue}
\hypersetup{
    colorlinks,
    citecolor=black,
    filecolor=black,
    linkcolor=black,
    urlcolor=black
}
\tableofcontents
}


\newpage

\section{Introduction}

There is a mounting evidence that
dynamical gravity can emerge 
along with space itself from non-gravitational 
quantum matter\cite{
Maldacena:1997re,
Witten:1998qj,
Gubser:1998bc,
PhysRevLett.96.181602,
2002PhLB..550..213K,
1126-6708-2007-07-062,
VanRaamsdonk:2010pw,
Lewkowycz2013,
2013JHEP...01..030K,
Headrick:2014aa,
Faulkner:2013aa,
Lashkari:2014aa,
Anninos:2017aa,
Lee:2013dln,
2014JHEP...03..051F,
PhysRevD.95.024031,
doi:10.1002/prop.201300020}.
What is lacking, though, is 
a concrete model 
from which a low-energy effective theory that
includes general relativity can be derived
from the first principle. 
The difficulty often lies in bridging the gap between
a microscopic model and the continuum limit.
If one starts with a discrete model, 
it is non-trivial to show the emergence of general relativity
in the continuum limit.
On the other hand, continuum theories usually require
new structures at short distances 
due to strong quantum fluctuations.   

A toy model of quantum gravity 
proposed in Ref. \cite{Lee:2020aa}
is well defined non-perturbatively
but simple enough that its continuum limit
can be understood in a controlled manner.
In the theory, the entirety of spacetime emerges 
as a collective behaviour of underlying quantum matter,
where the pattern of entanglement 
formed across local Hilbert spaces determines 
the dimension, topology and geometry of spacetime.
One unusual feature of the theory is that 
it has no pre-determined partitioning of the Hilbert space,
and the set of local Hilbert spaces
can be rotated within the total Hilbert space
under gauge transformations.
As a result, the theory has a large gauge group that
includes the usual diffeomorphism as a subset.
A  spacetime can be unambiguously determined from a state  
only after the Hilbert space decomposition is specified
in terms of some dynamical degrees of freedom as a reference.
As much as the entanglement is in the eye of the beholder,
the nature of emergent spacetime 
depends upon the Hilbert space decomposition.
Wildly different spacetimes with varying dimensions
and topologies can emerge out of one state, 
depending on what part of the total Hilbert space
 is deemed to comprise each local Hilbert space\cite{Lee:2021ta}.
With an arbitrary partitioning of the Hilbert space,
a generic state does not exhibit any local structure 
in the pattern of entanglement,
and the theory that emerges from the state is highly non-local.
This raises one crucial question :
{\it Is there a Hilbert space decomposition
that gives rise to a `local' theory of dynamical spacetime 
that includes general relativity?}  
In this paper, we address this question
by showing that 
there exists a natural partitioning of the Hilbert space
for states that exhibit local entanglement structures.
In a Hilbert space decomposition that supports 
a collection of  largely unentangled local clocks,
a massless graviton arises as a propagating mode
along with the local Lorentz invariance.

\section{A model of quantum gravity with emergent spacetime }
\label{sec:rev}

We begin with a brief review of the model introduced in Ref. \cite{Lee:2020aa}.
The fundamental degree of freedom is  an $M \times L$ real rectangular matrix
$\Phi^A_{~~i}$ with $1 \leq A \leq M$, $1 \leq i \leq L$
and $M > L$.
The row ($A$) labels flavours
and the column ($i$) labels sites.
The matrix can be viewed 
as representing a vector field with $M$ flavours
defined on a `space` with $L$ sites.
However, the dimension, topology and geometry
of the space is not pre-determined.
They will be determined from the pattern of
entanglement of states. 
The full kinematic Hilbert space 
is spanned by the set of basis states 
$ \cb \Phi \rb 
\equiv \otimes_{i,A} \cb \Phi^A_{~~i} \rb$,
where $ \cb \Phi^A_{~~i} \rb$ is the eigenstate of $\hat \Phi^A_{~~i}$. 
The conjugate momentum of $\hat \Phi$ is 
an $L \times M$ matrix $\hat \Pi^{i}_{~A}$.
The eigenstates of $\hat \Pi$ are denoted as
 $  \cb \Pi \rb \equiv \int d \Phi ~ e^{i  \Pi^{j}_{~A} \Phi^A_{~j} }  \cb \Phi \rb$.

The gauge symmetry is generated by
two operator-valued constraint matrices,
\bqa
\hat {\bf G}^i_{~j} & = & 
\frac{1}{2} \left( \hPiiA \hPhiAj + \hPhiAj \hPiiA 
+ i MC \delta^i_{~j}
\right), \nn
\hH^{ij} & =& \frac{1}{2} \left[
\left(  - \hPi \hPi^T + \frac{ \ta_0 }{M^2}  \hPi \hPi^T  \hPhi^T  \hPhi \hPi \hPi^T \right)^{ij} 
+ \left(  - \hPi \hPi^T + \frac{ \ta_0 }{M^2}  \hPi \hPi^T  \hPhi^T  \hPhi \hPi \hPi^T \right)^{ji} \right].
\eqa
Here, 
$(\hPi \hPi^T )^{ij} = \sum_A \hPi^i_{~A}  \hPi^j_{~A}$,
$(\hPhi^T \hPhi )_{ij} = \sum_A \hPhi^A_{~i}  \hPhi^A_{~j}$.
$C$ and $\tilde \alpha_0$ are constants.
${\bf G}$ is the generalized momentum constraint
that generates $\GLL$ transformation.
Under a transformation generated by the momentum constraint, 
$\Phi$ transforms as $\Phi \rightarrow \Phi g$, 
where $g \in \GLL$.
This includes permutations among sites, 
which can be viewed as a discrete version of the spatial diffeomorphism.
However, $\GLL$ is much bigger than the permutation group.
Under $\GLL$ transformations, the very notion of local sites can be changed 
because the local Hilbert space of $\Phi'=\Phi g$ at one site (column) 
is made of states that involve multiple sites (columns) of $\Phi$.
Therefore, there is no fixed notion of local sites in this theory.  
In Ref. \cite{Lee:2020aa}, only  $\SLL$ subgroup of $\GLL$ is taken 
as the gauge symmetry.
In this paper, we include the full $\GLL$ 
as the gauge group and 
introduce an additional parameter $C$
that controls gauge invariant Hilbert space\footnote{
One extra generator makes little difference in the large $L$ limit.}.
$\hH$ is the generalized Hamiltonian constraint,
which includes the Hamiltonian constraint of general relativity,
as will be shown later.
Both $\hat {\bf G}$ and $\hat H$ are invariant 
under the $O(M)$ flavour symmetry.  
The most general gauge transformation is generated by
$\hat \bG_{y_0} + \hat H_{v_0}$,
where
$\hat \bG_{y_0 }\equiv \tr{ \hat \bG y_0}$
and $\hat H_{v_0} \equiv \tr{ \hat H v_0}$.
$y_0$ is an $L \times L$ real matrix called shift tensor and
$v_0$ is an  $L \times L$ real symmetric matrix called lapse tensor.
The constraints, as quantum operators, satisfy the first-class algebra,
\bqa
\left[ \hat \bG_{x}, \hat \bG_{y} \right]  & = &     i \hG_{ ( yx-xy  ) }, \label{GGc} \\
\left[ \hat \bG_x, \hH_v \right]  & = &    i \hH_{  v x + x^T v  }, \label{GHc} \\
\left[ \hH_{v_1}, \hH_{v_2} \right]  & = &    
i \left[
 \hat C^{ijkl n}_{m}~ \hat \bG^m_{~n}
+ \frac{1}{M}\hat D^{ijkl }_{nm} ~ \hH^{mn}
\right] v_{1,ij} v_{2,kl},
\label{HHc}
\eqa
where
\bqa
\hat C^{ijkln}_{m} & = &
-4 \ta_0
\Bigl[
\hat U^{n(j} \hat U^{i)[l} \delta^{k]}_m - \hat U^{n [l} \hat U^{k](j} \delta^{i)}_m 
\Bigr]  \nn
&& + 4 \ta_0^2
 \Biggl[
-( \hU \hQ)^{(j}_{~[m} \hU^{\{ l [n} \hU^{n'] k \}} \delta^{i)}_{m']}
+ \hU^{ (j [n} ( \hQ \hU)_{[m'}^{~\{ k} \hU^{n'] i)} \delta^{l\}}_{m]} \nn
&& 
\hspace{1.2cm} 
+
 \hU^{ (j [n} ( \hU \hQ)^{\{ l}_{~[m} \hU^{n'] i)} \delta^{k\}}_{m']} 
- \hU^{ \{l [n} \hU^{n'] k\}} ( \hQ \hU)_{[m'}^{~( i}  \delta^{j)}_{m]} \nn
&&   \hspace{1.2cm} 
+  \frac{1}{M^2} \Bigl( 
M \hU^{(j [n} \hU^{n'] \{k} \delta^{l\}}_{[m} \delta^{i)}_{m']}
+ (M+2)  \hU^{(j [n} \hU^{\{l i) } \delta^{n']}_{[m} \delta^{k\}}_{m']}
+ 2 \hU^{(j [n} \hU^{n'] i)} \delta^{\{l}_{[m} \delta^{k\}}_{m']} \nn &&
   \hspace{1.2cm} 
- 2  \hU^{(j \{k} \hU^{l\} [n } \delta^{n']}_{[m} \delta^{i)}_{m']} 
- 2 \hU^{(j \{k} \hU^{[n' n]} \delta^{l\}}_{[m} \delta^{i)}_{m']}
- 2  \hU^{(ij) } \hU^{\{k [n } \delta^{n']}_{[m} \delta^{l\}}_{m']} 
\Bigr)
\Biggr] \delta^{m'}_{n'}, \nn
\hat D^{ijkl}_{nm} & = & 
-4i \ta_0 \left( \hat U^{kl} \delta^{ij}_{nm} - \hat U^{ij} \delta^{kl}_{nm}  \right)
\label{eq:ABC}
\eqa
with $\hat U = \frac{1}{M}( \hPi \hPi^T)$,
 $\hat Q = \frac{1}{M}(\hPhi^T \hPhi)$
 and
$\delta_{ij}^{kl} = \frac{1}{2}
\left(
   \delta_{i}^{k} \delta_{j}^{l} 
+ \delta_{i}^{l} \delta_{j}^{k}  \right)$\footnote{
Everywhere in Eqs. (\ref{GGc})-(\ref{eq:ABC}),
$\bG^i_{~j}$ can be replaced with its traceless counterpart
because $\hat C^{ijklm}_{m} =0$.
}. 
Pairs of indices
in 
$(i,j)$, 
$[n,n']$,
$[m,m']$,
$\{k,l\}$
are symmetrized.
The physical Hilbert space is spanned by 
gauge invariant states that satisfy
$\hat H_{v_0} \cb 0 \rb = 0$ and
$\hat G_{y_0} \cb 0 \rb = 0$
for any lapse tensor $v_0$ and shift tensor $y_0$.
Gauge invariant states are non-normalizable
with respect to the standard inner product 
for the scalars\cite{Lee:2020aa}.

Within the full Hilbert space,
we focus on a sub-Hilbert space 
that respects a specific flavour symmetry.
Here, we consider states that respect the
$O(N/2) \times O(N/2) \subset O(M)$ flavour symmetry,
where $N=M-L$.
The first $O(N/2)$ acts on flavours $A=L+1,..,L+N/2$
and the second $O(N/2)$ acts on flavours $A=L+N/2+1,..,M$.
The sub-Hilbert space can be spanned by the basis states labeled by
three matrix-valued collective variables,
\bqa
\cstrdual & = & 
\int D \Pi ~ e^{ - i  \left[
\sqrt{N} \sum_{a=1}^L 
  \Pi^{i}_{~a} 
q^a_{~i} 
+ 
\sum_{b=L+1}^{L+\frac{N}{2}} 
 P_{1,ij} 
\Pi_b^{~i} \Pi_b^{~j}
+
\sum_{c=L+\frac{N}{2}+1}^{M}
 P_{2,ij} 
\Pi_c^{~i} \Pi_c^{~j}
\right]
  }  
    \cb \Pi \rb,
\label{eq:symm}
\eqa
where $q$ is an $L \times L$ matrix
and $P_1$ and $P_2$ are $L \times L$ symmetric matrices.
Repeated indices $i,j$ are summed over all sites. 
Under an infinitesimal transformation generated by the constraints, 
$\cstrdual$ evolves as
\bqa
&& e^{ - i \epsilon ( \hat H_{v_0}  + \hat G_{y_0} ) } \cstrdual 
  = 
 \int 
 D q^{'} 
 D s^{'} 
DP^{'} 
DT^{'} 
 ~
 \cb q^{'}, P_1^{'}, P_2^{'} \rb \times \nn
  &&
 \hspace{3cm}
e^{
i N \epsilon \tr{
s^{'} \frac{ q^{'} - q }{\epsilon}
+  T_c^{'} \frac{ P_c^{'} - P_c }{\epsilon}
- 
 \cH[q,s^{'}, P_1, T_1^{'} , P_2, T_2^{'}  ] v
-  \cG[q,s^{'}, P_1, T_1^{'} , P_2, T_2^{'}  ]   y
}
}. 
\label{eq:epsilon_evolution}
\eqa
Here, 
$D q \equiv \prod_{i, a} d q^{a}_{~~i}$, 
 $D s \equiv \prod_{i, a} d s^{ i}_{~a}$, 
$D P \equiv \prod_{i \geq j } \left[ d P_{1,ij} dP_{2,ij} \right]$,
$D T \equiv \prod_{i \geq j } \left[ d T_{1}^{ij} dT_{2}^{ij} \right]$.
\eq{eq:epsilon_evolution} corresponds to the phase space
path integration representation 
of one infinitesimal step of evolution along a gauge orbit.
$\epsilon$ and $1/N$ play the role of an infinitesimal parameter time
and the Planck constant, respectively.
Identifying $ \frac{ q^{'} - q }{\epsilon}$ and
$\frac{ P_c^{'} - P_c }{\epsilon}$ as time derivatives 
of $q$ and $P_c$, respectively,
we conclude that
%
$s$ and $T_c$ are matrix-valued conjugate momenta of $q$ and $P_c$ with $c=1,2$, respectively.
The theory for the collective variables $\{ q,s,P_1,T_1, P_2, T_2 \}$ becomes
\bqa
S =  N \int d\tau ~ \tr{  
s \partial_\tau q
+T_1 \partial_\tau P_1
+T_2 \partial_\tau P_2
- \cH[q,s,P_1,T_1,P_2,T_2] v  
-  \cG[q,s,P_1,T_1,P_2,T_2] y  }, \nn
\label{eq:fullS}
\eqa
where
the generalized Hamiltonian and momentum 
matrices are given by
\bqa
 \cH[q,s,P_1,T_1,P_2,T_2]  = 
-U + \ta U Q U, ~~
\cG[q,s,P_1,T_1,P_2,T_2]  =  
  \left( s q  +  2 \sum_c  T_c P_c + 
  i \beta  I  \right),
 \label{HG}
 \eqa
where
$ U^{ij} = 
  \left( s s^T + T_1 + T_2 \right)^{ij}$,
$  Q_{ij}  =   
 \left( q^T q + \sum_c [ 4  P_c T_c P_c + i P_c ] \right)_{ij}$
 with
 $\beta =  \frac{M}{2N}(1+C) $,
$\ta = \frac{N^2}{M^2 - \ta_0 C M (L+1)}  \ta_0$,
$v= \left(   1 - \frac{ \ta_0 C  (L+1)}{M}  \right) v_0$,
$y= y_0   +  i \frac{ \ta_0  N}{M^2} 
[
(L+2) Uv_0 + \tr{Uv_0} I
]$. 
$I$ is the $L \times L$ identity matrix.
It is noted that 
$\ta$, $v$ and $y$ are renormalized 
by `contact' terms generated from normal ordering,
and \eq{HG} is exact for any $L$ and $N$.
From now on, we consider the large $N$ limit in which
we first take the large $L$ limit followed by the large $N$ limit
while tuning $\ta_0$  and $C$ such that $\ta \sim O(1)$ and $\beta = \frac{1}{2}$.

In Ref. \cite{Lee:2020aa}, 
the same Hamiltonian and momentum matrices
have been written for collective variables in the dual basis, $\cb \Phi \rb$.
That theory can be obtained
from \eq{HG} through a canonical transformation\footnote{
The advantage of using $\cb \Pi \rb$ basis to obtain
 the theory for  $\{ q, s, p_1, t_1, p_2, t_2 \}$ 
 is that it is easier to see the extra terms generated 
from the normal ordering
can be all absorbed into renormalization of 
$\ta$, $\beta$, $v$ and $y$.
},
\bqa
T_c = 4 t_c p_c t_c - i t_c, ~~~
P_c = \frac{1}{4 t_c}.
\eqa
In terms of $\{ q, s, p_1, t_1, p_2, t_2 \}$,
the generalized momentum constraint becomes
$\cG  =  
  \left( s q  +  2 \sum_c  t_c p_c -  \frac{i}{2}  I  \right)$.
The Hamiltonian takes the same form as \eq{HG} with
 \bqa
 U^{ij} = 
 \left( s s^T + \sum_c \left[ 4 t_c p_c  t_c - i t_c \right] \right)^{ij},
 ~~
Q_{ij}  =   
 \left( q^T q + p_1 + p_2  \right)_{ij}.
 \eqa
In the large $N$ limit, the collective variables $\{ q, s, p_1, t_1, p_2, t_2 \}$
become classical. 
The symplectic form defines the Poisson bracket,
$\{ A, B \} = 
\left(
    \frac{\partial A}{\partial q^\alpha_{~i } }  \frac{\partial B}{\partial s^i_{~\alpha} } 
-
\frac{\partial A}{\partial s^i_{~\alpha} }  \frac{\partial B}{\partial q^\alpha_{~i } }
      \right)
+ 
\delta^{kl}_{ij}
\left(
  \frac{\partial A}{\partial p_{c, ij}} \frac{\partial B}{\partial t_c^{kl} }
- \frac{\partial A}{\partial t_c^{kl} }  \frac{\partial B}{\partial p_{c,ij}}
\right)$.
From now on, we will use 
 $\{ q, s, p_1, t_1, p_2, t_2 \}$ 
 for describing emergent spacetime.
 There are $2L^2 + 4 \frac{L(L+1)}{2}$ phase space 
 degrees of freedom in these collective variables. 



\section{Frame and local clocks}

A semi-classical state with well-defined 
collective variables 
must satisfy the classical constraints,
\bqa
-U + \ta U Q U = 0, ~~~
s q  +  2 \sum_c  \td t_c p_c = 0,
%
\label{eq:HGzero}
\eqa 
where $\td t_c = t_c - \frac{i}{8 p_c}$.
This freezes $L(L+1)/2 + L^2$ collective variables
in terms of other variables\footnote{
For example, $s$ and $t_1$ can be solved in terms of the rest.}.  
The gauge redundancy 
removes  the same number of additional variables
from physical degrees of freedom,
leaving only $2L^2 + 4 \frac{L(L+1)}{2} - 2 \left( \frac{L(L+1)}{2} + L^2 \right)
= L (L+1)$ physical degrees of freedom.
Suppose we have an `initial' configuration of the collective variables
that satisfies \eq{eq:HGzero}.
A gauge orbit is generated by evolving the collective variables
with $\frac{\partial A}{\partial \tau} = \{ A, \cH_{v} + \cG_{y} \}$,
where $\tau$ is the parameter time.
The resulting equation of motion reads
\bqa
\partial_\tau \td t_c  =  -4 \td t_c v \td t_c   - \ta U v U  + \frac{1}{16} \frac{1}{p_c} v \frac{1}{p_c} - y \td t_c - \td t_c y^T, &&
\partial_\tau s  =  
-2 \ta U v U q^T - y s, \nn
\partial_\tau p_c  =   
4 p_c \td t_c v + 4 v \td t_c p_c   + p_c y + y^T p_c, &&
\partial_\tau q  =  
  2 s^T v  + q y.
\label{eq:EOM3}
\eqa
Different gauge orbits are obtained 
by evolving the initial collective variables with 
different lapse tensors ($v$) and shift tensors ($y$).
In general relativity,
different choices of lapse function and shift vector
only generate different spatial slices of one spacetime history.
In the present theory, spacetimes 
with different topologies and geometries
can be realized out of one state
with different choices of lapse and shift tensors\cite{Lee:2021ta}.
This is because in the present theory the set of gauge orbits 
is much larger than that of general relativity.
Each gauge orbit is labeled by the lapse tensor ($v_{ij}$) and the shift tensor
($y^i_{~j}$),
 which can be viewed as bi-local fields
defined on a space with $L$ sites.
In particular, the symmetric
rank $2$ lapse tensor has $L(L+1)/2$ independent entries 
while  the lapse function of general relativity,
being a scalar function, 
would have only $L$ independent parameters
for a system with $L$ sites.
The extra parameters in the lapse tensor are associated 
with the freedom of rotating the frame that defines local sites.
Under a frame rotation generated by 
$\GLL$,
$v$ transforms as $v \rightarrow g^T v g$ with 
$g \in$ $\GLL$.
Since one can always find $g 
\in O(L) \subset$ $\GLL$ 
in which the lapse tensor is diagonalized, 
the Hamiltonian with an off-diagonal lapse tensor can be viewed as 
the Hamiltonian with a diagonal lapse function in a rotated frame. 
Namely, $L$ eigenvalues of $v$ play the role of the lapse function defined on $L$ sites
while the rotation matrix that diagonalizes $v$ encodes the information
about the frame in which spatial sites are defined.

Therefore,
we need to choose a frame 
by fixing gauge
to extract a spacetime unambiguously.
As a first step, we impose a gauge fixing condition, 
\bqa
\cq \equiv q^T q=I.
\label{eq:qTq}
\eqa
With this, we demand sites are defined in a frame in which $q$ is orthonormal
as a matrix.
This still leaves the $O(L)$ subgroup of $\GLL$ unfixed.
One can fix the remaining $O(L)$ gauge symmetry
in terms of a variable that is used as local clocks.
For example, we can pick $p_1$ as our clock variable,
 choose a frame in which $p_1$ is diagonal,
and regard the $i$-th diagonal element of $p_1$  
as a physical time at site $i$.
With diagonal $p_1$, the local clocks 
are not entangled with each other. 

Let us now consider a state that has a local structure in a frame in which $\cq=I$ and $p_1$ is diagonal\footnote{
If $p_1$ has degenerate eigenvalues, 
there are multiple frames that diagonalize $p_1$.
In this case, we choose one of them.
}.
States with $d$-dimensional local structures
are the ones that are short-range entangled
(obeying the `area' law of entanglement)
when the sites are embedded in a 
$d$-dimensional manifold\cite{Lee:2020aa}.
In this case, we introduce a mapping from sites to a 
$d$-dimensional manifold
$ {\cal M}$ that has a well-defined topology, 
$r$ : $i \rightarrow r_i \in {\cal M}$,
where region $R_i \supset r_i$ is assigned to each site such that
$\cup_i R_i = {\cal M}$\cite{Lee:2020aa}.
For states with local structures, 
collective variables $t_c^{ij}$, $p_{c,ij}$
that are viewed as bi-local fields $t_c(r_i,r_j)$, $p_c(r_i,r_j)$
decay exponentially as functions of $r_i-r_j$ in ${\cal M}$.
To extract a spacetime in the gauge  
in which $\cq=I$ and $p_1$ is diagonal,
one should evolve the state
with the lapse and shift tensors 
that respect the gauge fixing conditions.
However, the shift and lapse tensors that keep $p_1$  
strictly diagonal are complicated\cite{Lee:2021ta}.
So, we take an alternative way of fixing gauge.
We still impose $\cq=I$, 
but relax the condition that $p_1$ is strictly diagonal.
Instead, we fix gauge by choosing simple lapse and shift tensors
such that local clocks remain `almost' unentangled 
under the Hamiltonian evolution.
The gauge orbits that respect the condition $\cq=I$
are generated by 
\bqa
\{
\cH_{v}+\cG_{- 2 \cs^T v }, \cG_{y'}
\},
\label{eq:residual}
\eqa
where 
$\cs \equiv s (q^{-1})^T$
and $y'$ is $L \times L$ real matrices that satisfy
$\cq y' + y^{'T} \cq = 0$.
$\cG_{y'}$ generates the unfixed $O(L)$ frame rotation. 
Within \eq{eq:residual},
we now choose a subset of constraints
that satisfy the following two conditions :
\bqa
(i) && \hspace{-0.4cm} \mbox{ the constraints in the subset satisfy the same algebra
that the momentum density} \nn
&& \hspace{-0.3cm}\mbox{and the Hamiltonian density
obey in general relativity}, \nn
(ii) && \hspace{-0.4cm} \mbox{
 the Hamiltonian density does not entangle initially
unentangled local clocks} \nn
&& \hspace{-0.3cm} \mbox{through an $O(L)$ frame rotation
in the limit that sites are weakly entangled.}
\label{conditions}
\eqa
These conditions are more explicitly explained as we
construct the momentum and Hamiltonian densities
in the following.
%

One can readily identify the momentum constraint of general relativity
from $\cG_{y'}$\cite{Lee:2020aa}.
Under an infinitesimal $\GLL$ transformation,
$\Phi^A_{~i}$ is transformed into
$\Phi^{'A}_{~i} = \Phi^{A}_{~j} (e^{-\epsilon y'})^j_{~i}$.
If  $\Phi^A_{i}$ varies slowly in the manifold, it
can be viewed as field $\Phi^A(r_i)$ defined on manifold ${\cal M}$, 
and the transformation can be written in the gradient expansion,
$ \Phi^{'A}(r_i)  = 
\left[ 1 - \epsilon \zeta(r_i) \right]  \hat \Phi^A(r_i) 
-\epsilon \xi^{\mu}(r_i)
\frac{\partial  \Phi^A(r_i)}{\partial r^{\mu}_i}  
- \epsilon 
\sum_{\bs=2}^\infty \frac{\xi^{\mu_1..\mu_\bs}(r_i)}{\bs!}
\frac{\partial^{\bs} 
 \Phi^A(r_i)}{\partial r^{\mu_1}_i..\partial r^{\mu_\bs}_i}  
$
with
$\zeta(r_i) =  \sum_j y^{'j}_{~i}$,
$\xi^\mu(r_i)  =  \sum_j y^{'j}_{~i} r^\mu_{ji}$,
$\xi^{\mu_1..\mu_\bs}(r_i)  =  \sum_j y^{'j}_{~i}
r^{\mu_1}_{ji}
..
r^{\mu_\bs}_{ji}
$
and
$r^\mu_{ji}=r^\mu_j - r^\mu_i$. 
Here $\zeta$ is the scale factor for the Weyl transformation.
$\xi^\mu$ is the shift vector 
and $\xi^{\mu_1..\mu_\bs}$
 with $\bs \geq 2$ 
corresponds to tensorial displacements for higher-derivative transformations.
One can single out the generator with each spin
by expressing $\cG$ in the gradient expansion,
$\cG_{y'} =  \int dr \Bigl(
\cD(r) \zeta(r) +  \cP_\mu(r)  \xi^\mu(r) + 
\sum_{\bs=2}^\infty 
\cP_{\mu_1..\mu_\bs}(r)  \xi^{\mu_1..\mu_{\bs}}(r) 
\Bigr)$.
Here, we use 
$\sum_i A_i = \int dr \tilde A(r)$ with $\tilde A(r_i) = V_i^{-1} A_i$
with $V_i$ denoting the coordinate volume of 
 region $R_i$ assigned to site $i$.
$\cD(r_i)  =  V_i^{-1} \cG^i_{~i}$,
$\cP_\mu(r_i)  =   V_i^{-1}  \left.  \frac{\partial \cG^i_{~j}}{\partial r^\mu_j} \right|_{j=i}$
and
$\cP_{\mu_1..\mu_\bs}(r_i)  =   \frac{V_i^{-1}}{\bs!}  \left.  
\frac{\partial^{\bs} \cG^i_{~j}}{\partial r^{\mu_1}_j..\partial r^{\mu_\bs}_j} 
\right|_{j=i}$
correspond to the generator of scale transformation,
the momentum density
and the generators of higher-derivative transformation, respectively.
The full algebra that $\cD$, $\cP_\mu$, $\cP_{\mu_1..\mu_\bs}$ satisfy is completely determined from \eq{HHc}.
In the absence of the tensorial displacement
($\xi^{\mu_1..\mu_\bs}=0$ for $\bs \geq 2$),
a simple closed algebra arises for $\cD$ and $\cP_\mu$,
\bqa
&&
 \left\{ 
\int dr \Bigl( \cD(r) \zeta_1(r) +  \cP_\mu(r)  \xi_1^\mu(r)  \Bigr), 
\int dr' \Bigl( \cD(r') \zeta_2(r') +  \cP_\nu(r')  \xi_2^\nu(r') \Bigr)
\right\} \nn
&& ~~~~~~~~~~
= \int dr 
\Bigl[ 
\left( \mathcal{L}_{\xi_1} \zeta_2(r) 
-\mathcal{L}_{\xi_2} \zeta_1(r) \right) \cD(r) 
+ 
\left( \mathcal{L}_{\xi_1} \xi_2 (r) \right)^\mu  \cP_\mu(r) 
\Bigr],
\label{eq:DPDP}
\eqa
where $ \mathcal{L}_{\xi}$ denotes the Lie derivative.
It is noted that $\cP_\mu$ indeed satisfies
the same algebra that the momentum density satisfies
in general relativity.

Identifying the Hamiltonian density of general relativity in \eq{eq:residual} 
is less straightforward 
because it can in general depend  on both $\cH$ and $\cG$.
As a candidate for the Hamiltonian,
we consider a constraint
that is labeled by a lapse function 
and written as a linear combination of $\cH$ and $\cG$,
\bqa
\bH_\theta \equiv \cH_{\theta} + \cG_{Y(\theta)}.
\label{eq:Hdensity}
\eqa
Here, 
$\theta_{ij} = \theta_i \delta_{ij}$ is the diagonal lapse tensor
 and $Y^i_{~j}(\theta)$ is a shift tensor that is linear in $\theta$.
$\theta_i$ is identified as the lapse function at position $r_i$,
and 
$\bH_\theta$
corresponds to the Hamiltonian associated with lapse function $\theta$.
For states with local structures, $\bH_\theta$ is written as $\int dr \bH(r) \theta(r)$,
where $ {\bf H}(r_i)  =  V_i^{-1}  \frac{\partial {\bf H}_\theta}{\partial \theta_i}$ is the Hamiltonian density.
In order for the gauge orbits to satisfy the gauge fixing condition in \eq{eq:qTq},
$Y(\theta)$ 
in \eq{eq:Hdensity}
should take the form of 
$Y(\theta)= - 2 \cs^T \theta  + \cq^{-1} ( F \theta - \theta F^T)$,
where $F_i^{~j}$ is a rank $2$ tensor
that in general depends on the collective variables.
The way $\bH_\theta$ is transformed under a scale transformation and a shift follows from \eq{HHc}\cite{Lee:2020aa}.
To the leading order in the derivative expansion
of the collective variables, 
the Poisson bracket between the momentum constraint and the Hamiltonian density is given by
\bqa
 \left\{
\int dr \Bigl( \cD(r) \zeta(r) +  \cP_\mu(r)  \xi^\mu(r) \Bigr), 
\int dr
  \theta(r)  \bH(r)
\right\}
 = 
\int dr ~
\left( 2 
\zeta(r) 
\theta(r) 
+  \mathcal{L}_{\xi}  \theta(r) \right)   \bH(r).
\label{eq:DPH}
\eqa
On the other hand, 
the Poisson bracket of two Hamiltonians can be written as
\bqa
 \{ \bH_{\theta_1}, \bH_{\theta_2} \} &&=
\bH_{ \theta_2 Y(\theta_1) + Y(\theta_1)^T \theta_2 - \theta_1 Y(\theta_2) - Y(\theta_2)^T \theta_1 } 
+ \{ Y(\theta_1)_{\cG}, Y(\theta_2)_{\cG} \}'
\nn
&& - \{ \cG_{Y(\theta_1)}, \cG_{Y(\theta_2)} \}'
+ \{ \cH_{\theta_1}, \cH_{\theta_2} \}'
+ \{ Y(\theta_1)_{\cG}, \cH_{\theta_2} \}'
+ \{ \cH_{\theta_1}, Y(\theta_2)_{\cG} \}'.
\label{eq:comH}
\eqa
Here, 
$\{ A_{x}, B_{y} \}'  \equiv 
\{ A^{i}_{~j}, B^{k}_{~l} \} x^{j}_{~i} y^{l}_{~k}$
denotes a reduced Poisson bracket,
where the derivatives of the Poisson bracket 
do not act on the variables in the subscripts of
$\{ A_{x}, B_{y} \}'$.
On the right hand side of \eq{eq:comH},  
the first term is proportional to ${\bf H}$ 
while the rest of the terms are all proportional to $\cG$.
The state obtained from 
an infinitesimal evolution with 
$\bH_{\theta_1}$
followed by an evolution with
$\bH_{\theta_2}$
must be related to the state obtained from
the sequence of evolutions performed in the opposite order
through a spatial diffeomorphism\cite{PhysRev.116.1322,TEITELBOIM1973542}. 
This implies that 
the term that is proportional to $\bH$ must vanish
in \eq{eq:comH}.
Therefore, 
the first requirement in (\ref{conditions}) leads to
\bqa
\theta_2 Y(\theta_1) + Y(\theta_1)^T \theta_2 - \theta_1 Y(\theta_2) - Y(\theta_2)^T \theta_1  = 0.
\label{eq:Hzerocondition}
\eqa
For $\cq = I$,
\eq{eq:Hzerocondition} is solved for $Y$ of the form,
$Y(\theta)= - 2 \cs^T \theta  + 
(  \Delta -2 \cs) \theta - \cq^{-1} \theta  (   \Delta -2 \cs^T ) \cq$,
where $\Delta$ is a symmetric matrix.
Here we consider $\Delta$ that is linear in $\cs$\footnote{
Different choices of $\Delta$ ultimately correspond
to different gauge fixing conditions.
}.
The only symmetric matrix linear in $\cs$ is
$\Delta =  2 \kappa ( \cs + \cs^T)$, 
where $\kappa$ is a parameter to be fixed
from the second requirement in (\ref{conditions}).
This gives
\bqa
Y(\theta) = \kappa ( - 2 \cq^{-1} \theta \cs \cq ) + ( 1-\kappa)
( -2 \cs^T \theta - 2 \cs \theta + 2 \cq^{-1} \theta \cs^T \cq ).
\eqa
Then, the Poisson bracket of $\bH_\theta$ becomes 
proportional to $\cG$,
\bqa
&& \{ \bH_{\theta_1}, \bH_{\theta_2} \} =
\theta_{1,i} \theta_{2,l} \bC^{iilln}_m \cG^m_{~n},
\label{eq:comH2}
\eqa
where
\bqa
&& C^{ijkln}_{m'}   (\cq^{-1})^{m'm}
  =  
4 \Biggl[
-\ta U^{nk} U^{li} \delta^{jm} 
 + \kappa \Bigl(
\ta U^{jk} U^{lm} \delta^{ni} + \cs^{kn} \cs^{jm} \delta^{li}
+ \cs^{li} \cs^{jm} \delta^{nk} - (\cs \cs^T)^{jk} \delta^{ni} \delta^{lm}
\Bigr) \nn
&&
+ (1-\kappa) \Bigl(
2\ta U^{nk} U^{li} \delta^{jm} 
 - \ta U^{jk}U^{lm} \delta^{ni}
 + \cs^{kn} \cs^{il} \delta^{jm}
 - \cs^{kn} \cs^{mj} \delta^{li} \nn &&
- \cs^{li} \cs^{mj} \delta^{nk}
- \cs^{kj} \cs^{ml} \delta^{ni}
+ \cs^{kj} \cs^{lm} \delta^{ni}
+ (\cs^2)^{kj} \delta^{ni} \delta^{lm}
 + \cs^{nk} \cs^{li} \delta^{jm} 
\Bigr) \nn
&&
+  \app^2 (\cs+\cs^T)^{in} (\cs+\cs^T)^{jk} \delta^{lm} 
-\app (\cs+\cs^T)^{in}
\left\{
\app \cs^{ml} - \kappa  \cs^{lm}
\right\} \delta^{jk}
%
\nn
&&
+ 
\Bigl(
\app^2 \cs^{kj} \cs^{ml} 
+ \bpp^2 \cs^{jk} \cs^{lm}
- \bpp \app \left\{ \cs^{kj} \cs^{lm} + \cs^{jk} \cs^{ml} \right\}
\Bigr)
\delta^{ni} 
\nn &&
- \Bigl(
\app^2  \big\{ (\cs^2)^{kj} +  (\cs^T \cs)^{jk} \big\}
- \app \bpp \big\{ (\cs \cs^T)^{jk}+  (\cs^2)^{jk} \big\}
\Bigr)
 \delta^{ni} \delta^{lm}
\Biggr]
 - \Bigl[ (ij) \leftrightarrow (kl) \Bigr]. 
\label{eq:Cijklnm}
\eqa
In \eq{eq:Cijklnm},
we use the constraint $\cG=0$ 
and the gauge fixing condition $\cq=I$
to simplify the expression
for $C^{ijkln}_{m}$.
%
For the state that has a local structure,
\eq{eq:comH2} can be written as
 \bqa
&& \Bigl\{  \int dr ~\theta_1(r) \cH(r) ,  \int dr ~\theta_2(r) \cH(r)  \Bigr\}  
=  
\int dr ~
  \left( \theta _1 \nabla_{\mu_1} \theta _2  - \theta _2 \nabla_{\mu_1} \theta _1 \right) \times \nn
  &&~~~~~~
\Bigl[
F^{\mu_1}(r)   \cD(r)  
+
G^{\mu_1 \mu_2}  \cP_{\mu_2}(r) 
 +\sum_{\bs =3}^\infty
G^{ \mu_1 \mu_2..\mu_{\bs}}
    \cP_{\mu_2..\mu_{\bs}}(r)
\Bigr], 
\label{eq:HH}
 \eqa 
to the leading order in the derivative of the lapse function,
where
\bqa
&& F^\mu(r)  =   \left. \frac{1}{2}  \sum_{i,l,m,n} C^{ i i ll n }_{m}  r^\mu_{li} \right|_{\frac{r_n+r_m}{2} = r}, ~~
G^{\mu_1 \mu_2}
=
 \left.
\frac{1}{2} \sum_{i,l,m,n} C^{iilln}_m
   r^{\mu_1}_{li}   r^{\mu_2}_{nm}
   \right|_{\frac{r_n+r_m}{2} = r}, \nn
&& G^{\mu_1 \mu_2.. \mu_{\bs}}
=
 \left.
\frac{1}{2} \sum_{i,l,m,n} C^{iilln}_m
  r^{\mu_1}_{li} 
r^{\mu_2}_{nm} ..
r^{\mu_{\bs}}_{nm} 
   \right|_{\frac{r_n+r_m}{2} = r}
  \label{eq:sgmunu} 
\eqa
with 
the sum over $m$ and $n$ restricted to 
those sites for which  $\frac{r_n+r_m}{2} = r$.
If the third and higher moments of $C^{iilln}_m$ are small,
the spin-${\bs}$ field $G^{\mu_1 \mu_2..\mu_{\bs}}$ is negligible 
for $\bs \geq 3$.
In this limit, $\cD(r)$, $\cP_\mu(r)$ and $\cH(r)$
form a closed algebra to the leading order 
in the derivative expansion.
In particular, Eqs. (\ref{eq:DPDP}), (\ref{eq:DPH}) and (\ref{eq:HH})
restore the algebra that the momentum density and the Hamiltonian density satisfy
in general relativity\cite{PhysRev.116.1322,TEITELBOIM1973542} 
provided that 
$\frac{G^{\mu \nu}+G^{\nu \mu}}{2}$ is identified as $-\cS g^{\mu \nu}$,
where $\cS$ and $g^{\mu \nu}$
are the signature of time
and the space metric, respectively,
in the convention in which the spatial metric is positive.

Interestingly, the metric depends on $\kappa$
that  parameterizes the $O(L)$ frame rotation 
included in the Hamiltonian.
The value of $\kappa$ affects 
how dynamical variables 
run under the Hamiltonian evolution
because the very notion of local sites
is rotated under the frame rotation.
To fix $\kappa$, we turn to the second condition 
in (\ref{conditions}).
%
For this, we consider 
the simplest `pre-gometric' state in which 
collective variables are ultra-local with no inter-site entanglement,
\bqa
p_1 = p_2 = \cq = I.
\label{eq:ultralocal}
\eqa
The rate at which the clock variable 
$p_1$ runs under the Hamiltonian evolution
depends on the conjugate momentum $\td t_1$.
On the other hand,
$\td t_1$ along with $\td t_2$ is subject to 
the gauge constrains in \eq{eq:HGzero},
\bqa
\cs  + 2 (\td t_1 + \td t_2)  =  0, ~~~~
8 \td t_1^2 + 8 \td t_2^2 + 4 \td t_1 \td t_2 +  4 \td t_2 \td t_1 + \frac{1}{8}-\frac{1}{3 \ta}  =  0. 
\eqa
For the stationary clock with $\td t_1=0$,
$\td t_2$ and $\cs$ are determined to be
$ \td t_2 =  -\frac{\cs}{2} = \bar t I$ with
 $\bar t = \left[ \frac{1}{8} \left( \frac{1}{3 \ta}  -\frac{1}{8} \right) \right]^{1/2}$.
Now, consider a small perturbation to the conjugate momentum of the clock variable
$\td t_1 =  t' $ with $|t^{' ij}| \ll \bar t$.
The constraints determine $\td t_2$ and $\cs$ to be 
$\td t_2 = \bar t - \frac{t'}{2}$ 
and
$\cs = -2 \bar t - t'$
to the linear order in $t'$.
Under the evolution generated by $\bH_\theta$, the clock variable evolve as
\bqa
\partial_\tau p_1 = p_1 R + R^T p_1,
\eqa
where $R = 4 t' \theta + 
 \kappa ( - 2 \theta \cs  ) + ( 1-\kappa)
( -2 \cs^T \theta - 2 \cs \theta + 2 \theta \cs^T  )$.
The anti-symmetric part of $R$,
which generates $O(L)$ rotation of the clock variable,
is given by $R-R^T = 2(5-4 \kappa) ( t' \theta - \theta t')$.
The $O(L)$ rotation, if present, would mix the clock at one site with the one at another site. 
We choose $\kappa$ such that local clocks 
do not get entangled through $O(L)$ rotation
in the limit that the sites are weakly entangled.
This leads to 
\bqa
\kappa = \frac{5}{4}.
\label{eq:kappa}
\eqa
With this, the Hamiltonian density is uniquely fixed.
We now examine the dynamics of 
the spacetime that emerges from a state 
with a three-dimensional local structure
and study the spin-$2$ mode that propagates
on top of the semi-classical background spacetime.

\section{Background spacetime}

Since the metric determined from 
\eq{eq:Cijklnm} 
and 
\eq{eq:sgmunu} 
depends only on $\cs$ and $U$,
it suffices to understand the evolution 
of $U$ and $\cs$ to understand the dynamics of geometry.
We choose the lapse $\theta=I$
which corresponds to the uniform lapse function $\theta(r)=1$.
The equations of motion for $U$ and $\cs$ are given by
\bqa
\frac{ \partial U}{\partial \tau} = 2 \cs U + 2 U \cs^T, ~~~~
\frac{ \partial \cs}{\partial \tau} = -2 \ta U^2 + 2 \cs \cs^T.
\label{eq:dtau us}
\eqa
While $U$ is a symmetric matrix, $\cs$ is a general $L \times L$ matrix.
However, we can focus on the sub-Hilbert space 
in which $\cs$ is symmetric 
because \eq{eq:dtau us} preserves the symmetric nature 
of initial $\cs$.

 \begin{figure}[ht]
 \begin{center}
   \subfigure[]{
 \includegraphics[scale=0.8]{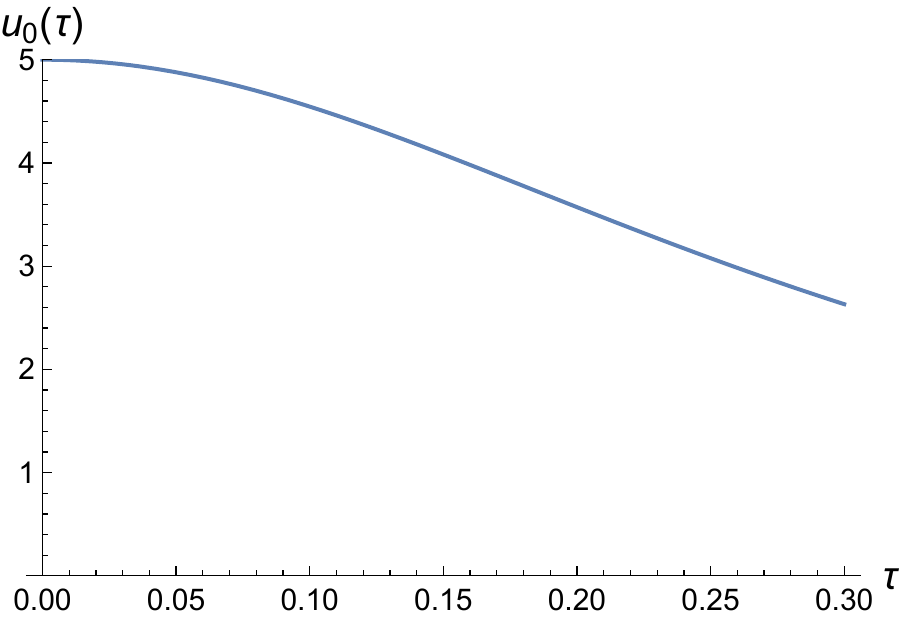} 
  \label{fig:u0}
 } 
  \hfill
 \subfigure[]{
 \includegraphics[scale=0.8]{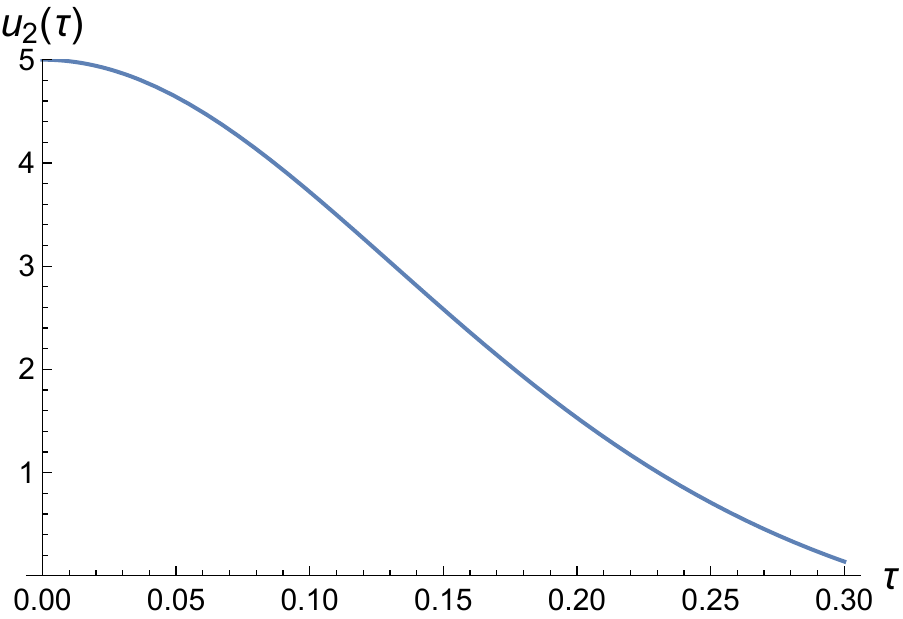} 
 \label{fig:u2}
 } 
    \subfigure[]{
 \includegraphics[scale=0.8]{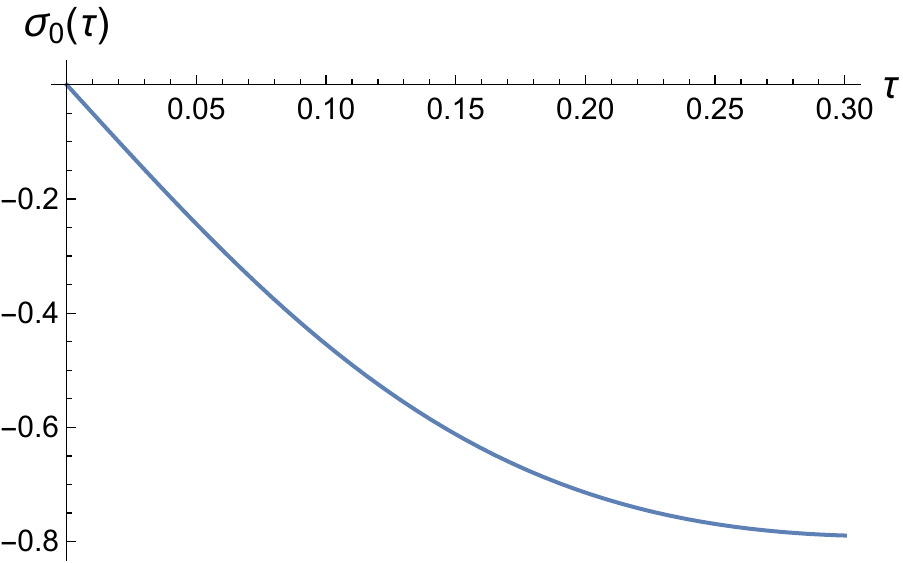} 
  \label{fig:s0}
 } 
  \hfill
 \subfigure[]{
 \includegraphics[scale=0.8]{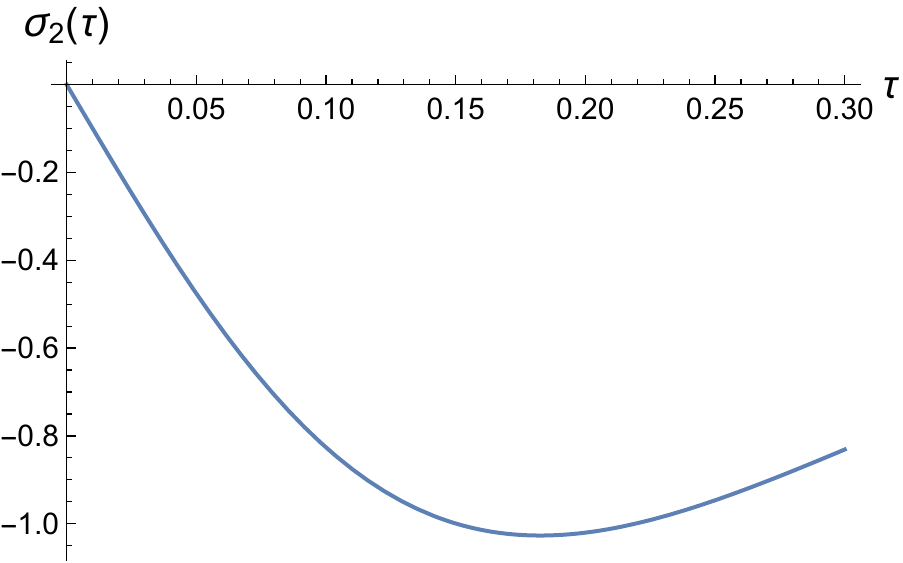} 
 \label{fig:s2}
 } 
  \end{center}
 \caption{
 $u_0(\tau)$,
 $u_2(\tau)$,
$\sigma_0(\tau)$,
$\sigma_2(\tau)$
plotted as functions of $\tau$
for 
$\ta = 0.1$,
$\bpp=5/4$,
$\bar u_0=\bar u_2=5$
and $\bar \sigma_0 = \bar \sigma_2=0$. 
 }
 \label{fig:us}
 \end{figure}

Let us consider a state with a three-dimensional local structure with 
$T^3$ topology, the translational and space inversion symmetry.
For simplicity, let us consider $L=\ell^3$ for an integer $\ell$.
The natural mapping from sites to $T^3$ is 
$r_i = 
\left( 
i \mod \ell, 
~~\lfloor \frac{i}{\ell} \rfloor \mod \ell , 
~~\lfloor \frac{i}{\ell^2} \rfloor \mod \ell 
\right)$
for $1 \leq i \leq L$,
where $\lfloor x \rfloor$ is the floor function.
In $T^3$, the periodic boundary condition is used
with $(x,y,z) \sim (x+\ell,y,z) \sim (x,y+\ell,z) \sim (x,y,z+\ell)$.
In the Fourier space,
the collective variables satisfy
\bqa
\frac{ \partial U_k}{\partial \tau} = 4 \cs_k U_k, ~~~~
\frac{ \partial \cs_k}{\partial \tau}= -2 \ta U_k^2 + 2 \cs_k^2,
\label{eq:EOMuniform}
\eqa
where
$k = \frac{2\pi}{\ell} ( n_1, n_2, n_3)$ with $-\ell/2 \leq n_i < \ell/2$ denotes
three-dimensional momenta,
$ U_k   =  \sum_{j} e^{ -i k r_j } ~ U_{r_j, 0}$ and
$\cs_k  =  \sum_{j} e^{ -i k r_j } ~ \cs^{r_j, 0}$.
The solution of \eq{eq:dtau us} 
is given by
\bqa
U_k(\tau)  =  \frac{U_k(0)}{(1-2 \cs_k(0) \tau)^2
+4 \ta   U_k(0)^2  \tau^2 }, ~~~
\cs_k(\tau)  = 
 \frac{
 \cs_k(0)
-2 \left[ \cs_k(0)^2 
+\ta  U_k(0)^2
\right] \tau }{(1-2
   \cs_k(0) \tau)^2+4 \ta  U_k(0)^2  \tau^2}.
 \eqa
 For states with local structures, 
 $U_k$ and $\cs_k$
 are analytic functions of $k$ and can be expanded around $k=0$ as
 \bqa
 U_k(\tau)  =  u_0(\tau)  + u_2(\tau)  k^2 + O(k^4), ~~~
 \cs_k(\tau)   = \sigma_0(\tau)  + \sigma_2(\tau)  k^2 + O(k^4),
 \label{eq:initialuniform}
 \eqa
 where $k^2 \equiv \sum_{\mu=1}^3 (k_\mu)^2$.
 While $U_k$ and $\cs_k$  have only the discrete rotational symmetry
 at the lattice scale,
 at small $k$ the full rotational symmetry emerges. 
The coefficients of $U_k$ and $\cs_k$ evolve as 
\bqa
&& u_0(\tau) =
\frac{\bar u_0}{ (1-2 \bar \sigma_0 \tau)^2+ 4 \ta  \bar u_0^2 \tau^2 } ,~~
u_2(\tau) = \frac{\bar u_2 - 4 \tau \left[\bar u_2 \left(\bar \sigma_0 - \bar \sigma_0^2 \tau +\ta  \tau \bar u_0^2\right)+
   \bar \sigma_2 \bar u_0 (2 \bar \sigma_0 \tau-1) \right] }{\left[(1-2\bar \sigma_0 \tau)^2+4 \ta \bar u_0^2  \tau^2  \right]^2}, \nn 
&& \sigma_0(\tau) =\frac{-2\bar \sigma_0^2 \tau+\bar \sigma_0-2 \ta  \bar u_0^2  \tau}{(1-2
  \bar \sigma_0 \tau)^2+4 \ta \bar u_0^2  \tau^2}, ~~
\sigma_2(\tau) = \frac{\bar \sigma_2 (1-2\bar \sigma_0 \tau)^2-4 \ta   \bar u_0 (-2
  \bar \sigma_0  \bar u_2 \tau +\bar \sigma_2 \bar u_0  \tau +\bar u_2) \tau}{\left[(1-2\bar \sigma_0 \tau)^2+4 \ta  
  \bar u_0^2 \tau^2 \right]^2}, \nn
\eqa
where 
$\bar u_0 = u_0(0)$, $\bar u_2 = u_2(0)$, 
$\bar \cs_0 = \cs_0(0)$ and $\bar \cs_2 = \cs_2(0)$. 
These functions are plotted in  \fig{fig:us}
for a choice of initial condition.

 \begin{figure}[ht]
 \begin{center}
     \subfigure[]{
 \includegraphics[scale=0.8]{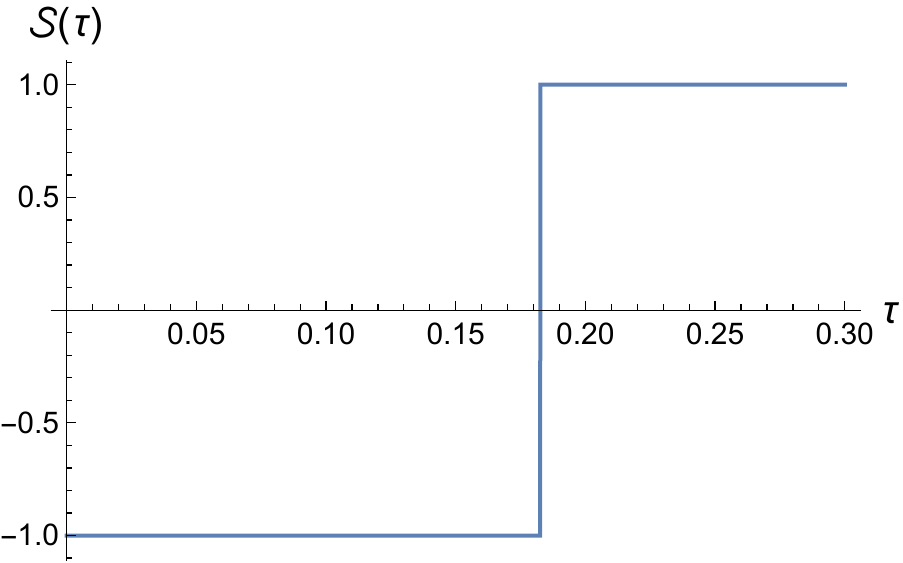} 
  \label{fig:sig}
 } 
  \hfill
 \subfigure[]{
 \includegraphics[scale=0.8]{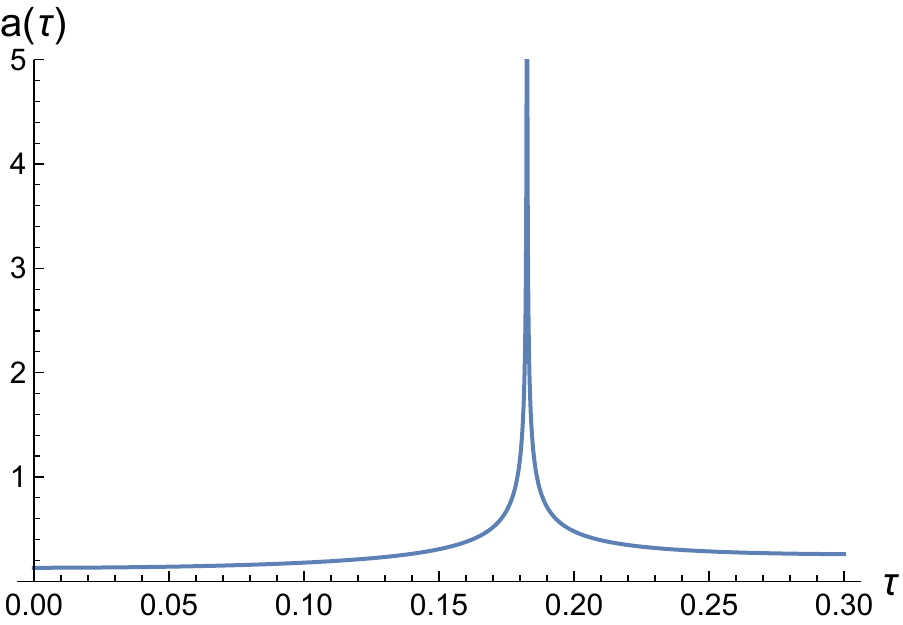} 
 \label{fig:a}
 } 
  \end{center}
 \caption{
The signature and the scale factor
plotted as functions of $\tau$ for 
the same parameters used in \fig{fig:us}.
Initially, the de Sitter-like spacetime 
with the Lorentzian signature is realized,
where the scale factor increases with increasing parameter time.
The spacetime undergoes a phase transition
into a Euclidean spacetime around $ \tau_c = 0.182$. 
The signature-changing phase transition is accompanied
with the divergent scale factor\cite{Lee:2020aa}.
 }
 \label{fig:sa}
 \end{figure}

The signature and the spatial metric is determined from
 \eq{eq:sgmunu}, which reduces to 
  \bqa
-\cS g^{\mu \nu}(r)
=
 \left.
-6
 \sum_{m,n} 
 \left[
\ta U^2 - \cs^2 \right]^{nm} r^\mu_{nm}  r^\nu_{nm}
 \right|_{\frac{r_n+r_m}{2} = r}.
\label{eq:sgmunu0}
\eqa
Because the spatial metric gives the uniform and flat three torus
with a time dependent scale factor,
we obtain the Friedmann–Robertson–Walker (FRW) metric\cite{Lee:2020aa},
\bqa
ds^2 = \cS(\tau) d\tau^2 + a(\tau)^2 dx^\mu dx^\mu, 
\label{eq:ds2}
\eqa
where $\cS(\tau) $ is the signature of time and 
$a(\tau)$ is the scale factor of the uniform space given by
\bqa
\cS(\tau) &= & -\sgn{\left[  \ta u_0(\tau) u_2(\tau) - \sigma_0(\tau) \sigma_2(\tau)   \right]}, \nn
a(\tau) &= &  
\frac{1}{
 \sqrt{24 
| \ta u_0(\tau) u_2(\tau) - \sigma_0(\tau) \sigma_2(\tau)  |
}}
.
\label{eq:Sa}
\eqa
The signature and scale factor associated
with the solution shown in \fig{fig:us}
are plotted in \fig{fig:sa}.
The saddle point solution determines the background spacetime. 
In the following, we examine the dynamics of the spin-$2$ mode 
that propagates on this spacetime.
At a critical parameter time $\tau_c \approx 0.182$,
there is a phase transition at which 
the scale factor diverges and the signature of spacetime jumps.
Signature-changing transitions have been also studied in Ref. 
\cite{PhysRevD.95.124014,PhysRevD.98.086015}.
Here we will focus on the range of parameter time
($0< \tau < \tau_c$)
in which the spacetime is Lorentzian ($\cS=-1$)
and the space is expanding.

\section{Graviton}

Small fluctuations of the collective variables above the translationally invariant solution are described by
the linearized equations,
\bqa
\frac{ \partial \delta U_{k_1 k_2}}{\partial \tau} &=& 2 (\cs_{k_1} + \cs_{k_2}) \delta U_{k_1 k_2} + 2 (U_{k_1}+U_{k_2}) \delta \cs_{k_1 k_2}, \nn
\frac{ \partial \delta \cs_{k_1 k_2}}{\partial \tau} &=& -2 \ta (U_{k_1}+U_{k_2}) \delta U_{k_1 k_2}  + 2 (\cs_{k_1}+\cs_{k_2}) \delta \cs_{k_1 k_2},
\label{eq:dtaudeltaUs}
\eqa
where
$ \delta U_{k_1 k_2}   =  \sum_{r_1 r_2} e^{ -i k_1 r_1 -i k_2 r_2 } ~ \delta U_{r_1, r_2}$ and
$ \delta \cs_{k_1 k_2}   =  \sum_{r_1 r_2} e^{ -i k_1 r_1 -i k_2 r_2 } ~ \delta \cs^{r_1, r_2}$.
A deviation of $g^{\mu \nu}$ denoted as $h^{\mu \nu}$ 
is linearly related to $\delta U$ and $\delta \cs$.
In the Fourier space, the metric fluctuation with momentum $k$
is written as 
\bqa
h^{\mu \nu}_k &=& -
6 \cS \Biggl\{
\Delta^{\mu \nu} 
\Bigl[
\ta ( U_{k_1} + U_{k_2} ) \delta U_{k_1 k_2}
-  ( \cs_{k_1} + \cs_{k_2} ) \delta \cs_{k_1 k_2}
\Bigr] \Biggr\}_{k_1=k_2=\frac{k}{2}},
\label{eq:hmunu}
\eqa
where $\Delta^{\mu \nu} \equiv
\left( \frac{\partial}{\partial k_{1\mu}} - \frac{\partial}{\partial k_{2\mu}} \right)
\left( \frac{\partial}{\partial k_{1\nu}} - \frac{\partial}{\partial k_{2\nu}} \right)$.
A traceless transverse mode can be isolated as
$h_k \equiv a^2 \epsilon_{\mu \nu} h^{\mu \nu}_k$,
where $\epsilon_{\mu \nu}$ is a time-independent polarization tensor
that satisfies
$\epsilon_{\mu \nu} k^{\nu} = 0$
and $\epsilon^{\mu}_{~\mu}=0$.
Due to $\epsilon_{\mu \nu} 
\frac{\partial}{\partial k_\mu} U_{k}=
\epsilon_{\mu \nu} \frac{\partial^2}{\partial k_\mu \partial k_\nu} U_{k}=0$,
which is guaranteed by the inversion symmetry 
and the discrete rotational symmetry of the background configuration,
the traceless transverse mode is given by
\bqa
h_k =-
12 \cS a^2  
\epsilon_{\mu \nu} \Biggl\{
\ta  U_{k/2} \Delta^{\mu \nu}  \delta U_{k_1 k_2}
-    \cs_{k/2} \Delta^{\mu \nu} \delta \cs_{k_1 k_2}
\Biggr\}_{k_1=k_2=\frac{k}{2}}.
\eqa
The equation of motion of $h_k$
directly follows from \eq{eq:dtaudeltaUs}.
To the second order in $k$
and the number of derivatives in time,
it becomes 
\bqa
\ddot{ h }_k
+
\left[ \frac{15 u_0 \dot{u_2} -u_2 \dot{u_0}  }{4 u_0 u_2} \right]
 \dot{  h}_k
  - \cS k^2 h_k=0.
\label{eq:gravitonEOM}
\eqa
Here, 
$\dot{f} \equiv \partial_\eta f$ and
$\ddot{f} \equiv \partial_\eta^2 f$,
where $\eta$ is the conformal time defined from $d \eta = a(\tau)^{-1} d\tau$.
\eq{eq:gravitonEOM} describes a massless spin-$2$ mode
propagating 
in the presence of time dependent 
background metric and other fields\cite{Lifshitz:2017vu}.
In the Lorentzian spacetime ($\cS = -1$),
the low-energy graviton propagates with speed $1$
in the background metric given 
by Eqs. (\ref{eq:ds2}) and (\ref{eq:Sa}).
This indicates that the local Lorentz invariance emerges 
in the frame that supports local clocks\footnote{
If we choose $\kappa$ which is different from \eq{eq:kappa},
the mode becomes either sub-luminal or super-luminal.
This is a manifestation of the fact that
the spacetime and the effective theory theory
that describes excitation above the spacetime 
take different forms once the total Hilbert space is partitioned
into local Hilbert spaces differently.}.
The uniqueness of general relativity
as a Lorentz-invariant interacting theory of gapless spin-$2$ particle\cite{weinberg_1995} suggests
that the present theory  includes
general relativity as an effective theory 
for states with local structures
in a gauge that supports
an extended space and local clocks.

\section{Toward an isolated graviton}

One way to understand why the gapless graviton is present 
as a propagating mode
is to view the theory for the collective variables in \eq{eq:fullS} 
as the holographic dual of a boundary theory.
In this perspective,
the exponent of the wavefunction in \eq{eq:symm} is identified as
the action of a non-unitary boundary theory, 
and the Hamiltonian constraint becomes the generator 
of the evolution along the emergent radial direction\cite{Lee:2013dln,Lee2016}.
As is the case for the AdS/CFT correspondencer\cite{
Maldacena:1997re,
Witten:1998qj,
Gubser:1998bc}, 
every global symmetry of the boundary theory is promoted 
to a gauge symmetry in the bulk,
and an unbroken symmetry in the boundary 
gives rise to a gapless gauge field in the bulk\cite{doi:10.1142/S0217751X13501662,Bednik_thesis}.  
In the present theory, the gauge symmetry includes the space diffeomorphism
generated by the momentum constraint.
Therefore, the unbroken translational symmetry in \eq{eq:symm}
gives rise to a gapless  gauge field associated with it\footnote{
Strictly speaking, 
there is only the discrete translational symmetry 
due to the lattice structure in the boundary theory.
Consequently, the momentum conservation can be violated
by the reciprocal momentum which is
inversely proportional to the lattice spacing.
However, the full momentum conservation emerges
in the long-wavelength limit.
This is because one can not construct an operator 
that carries the reciprocal momentum out of a finite
number of modes whose momenta are taken to be 
increasingly small in the long-wavelength limit. 
}, which is the gapless graviton.

Besides the gapless graviton, 
there also exists a continuum of spin-$2$ modes 
in the present model.
Those modes are labeled by the relative 
momentum of bi-local fields,
\bqa
h^{\mu \nu}_{k,q} &=& -
6 \cS \Biggl\{
\Delta^{\mu \nu} 
\Bigl[
\ta ( U_{k_1} + U_{k_2} ) \delta U_{k_1 k_2}
-  ( \cs_{k_1} + \cs_{k_2} ) \delta \cs_{k_1 k_2}
\Bigr] \Biggr\}_{
k_1=\frac{k+q}{2},
k_2=\frac{k-q}{2}
},
\label{eq:hmunuq}
\eqa
where 
$k$ is the center of mass momentum
and
$q$ is the relative momentum of $\delta U_{k1, k2}$.
The gapless graviton in 
\eq{eq:hmunu}
corresponds to the mode with $q=0$.
If both $k$ and $q$ are transverse to the polarization 
($\epsilon_{\mu \nu} k^{\nu} = \epsilon_{\mu \nu} q^{\nu} = 0$),
$h_{k,q} \equiv a^2 \epsilon_{\mu \nu} h^{\mu \nu}_{k,q}$
satisfies the equation of motion similar to \eq{eq:gravitonEOM},
\bqa
\ddot{ h }_{k,q}
+
\left[ \frac{15 u_0 \dot{u_2} -u_2 \dot{u_0}  }{4 u_0 u_2} \right]
 \dot{  h}_{k,q}
  - \cS (k^2+m_q^2) h_{k,q}=0,
\label{eq:gravitonEOMq}
\eqa
where the $q$-dependent mass 
goes as $m_q^2 = q^2$ in the small $q$ limit. 
The existence of the continuum of modes
is a consequence of the fact that 
both the center of mass momentum 
and the relative momentum of the bi-local fields are conserved.
This feature is shared with 
the holographic descriptions of vector models
in the large $N$ limit
\cite{2002PhLB..550..213K,
Das:2003vw,
Koch:2010cy,
Douglas:2010rc,
2013arXiv1303.6641P,
Leigh:2014tza,
2015PhRvD..91b6002L,
2014arXiv1411.3151M,
Vasiliev:1995dn,
Vasiliev:1999ba,
Giombi:2009wh,
Vasiliev:2003ev,
Maldacena:2011jn,
Maldacena:2012sf,
2014PhRvD..90h5003S,
Lee2016}.
In order to remove this unrealistic feature,
one has to allow mixing between
modes with different relative momenta.
In this section, 
we discuss how such mixing 
arises through $1/N$ corrections.

 \begin{figure}[ht]
 \begin{center}
 \includegraphics[scale=1]{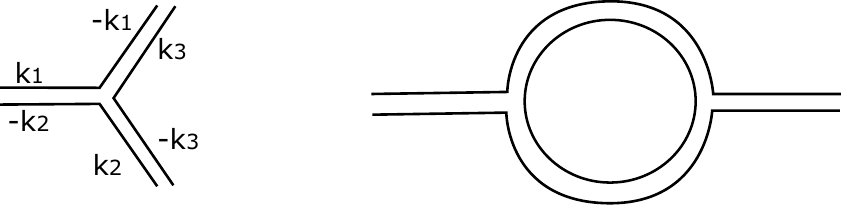} 
  \end{center}
 \caption{
 The left diagram shows the cubic vertex for the bi-local field $\delta T_2$ 
 shown  in \eq{eq:cubicV}. 
 The momentum along each single line is preserved.
 As a result, the one-loop self-energy, shown in the right, is diagonal
 both in the center of mass momentum and the relative momentum
 of the bi-local field.
 }
 \label{fig:cubic0}
 \end{figure}
 
To consider $1/N$ corrections,
we need the full theory in \eq{eq:fullS}.
The theory for the propagating modes
 can be obtained by expanding 
the collective fields around the the saddle-point 
$\{ \bar q, \bar s,  \bar P_1,\bar T_1, \bar P_2, \bar T_2  \}$ 
and writing down the theory
for the fluctuating variables 
$\{ \delta q, \delta s,  \delta P_1,\delta T_1, \delta P_2, \delta T_2  \}$.
The quadratic part determines the free propagator,
which can be obtained from the equation of motion
obeyed by the fluctuating variables.
The full theory also include interaction vertices.
For example, 
$\ta U Q U$ in \eq{eq:fullS}
includes a cubic vertex for $\delta T_2$,
$ 4 \ta \tr{ \delta T_2 \bar P_2 \delta T_2 \bar  P_2 \delta T_2 }$
for the choice of $v=I$.
In the Fourier space, the vertex can be written as
\bqa
4 \ta \int dk_1 dk_2 dk_3 ~
V_{k_1,k_2,k_3}
~
\delta T_{2}^{k_1,-k_2} \delta T_{2}^{k_2,-k_3} \delta T_{2}^{k_3,-k_1},
\label{eq:cubicV}
\eqa
where 
$V_{k_1,k_2,k_3} =
\bar P_{2; k_2,-k_2} \bar  P_{2; k_3,-k_3}$.
At the saddle-point, the collective fields have non-zero expectation values 
only for the modes with zero center of mass momentum
due to the translational invariance.
In general, loop corrections can modify the quadratic action 
for $\delta T_2$ as
\bqa
\delta S  = \int dk_1 dk_2 dk_3 dk_4 ~
\Sigma_{k_1,k_2; k_3, k_4} 
\left( \delta T_2^{k1,k2} \right)^*
\delta T_2^{k3,k4},
\eqa
where $\Sigma_{k_1,k_2; k_3, k_4} $ denotes the self-energy
of $\delta T_2$,
which is suppressed by $1/N$ compared to \eq{eq:fullS}.
The self-energy generated from 
\eq{eq:cubicV} through the one-loop diagram in  \fig{fig:cubic0}
takes the form of
\bqa
\Sigma_{k_1,k_2; k_3,k_4} \sim 
\left(
\delta_{k_1,k_3} \delta_{k_2,k_4}
+ \delta_{k_1,k_4} \delta_{k_2,k_3}
\right)
\int dq 
V_{k_1,-k_2,-q}
V_{-k_1,q,k_2}
G_2(-k_1,-q) G_2(q,-k_2),   
\eqa 
where $G_2(k_1,k_2)$ is the propagator of $\delta T_{2}^{k_1,k_2}$.
It is noted that the self-energy 
is still diagonal both in the center of mass momentum
and the relative momentum\footnote{
$\delta T_2^{k_1 k_2} = \delta T_2^{k_2 k_1}$ because 
the bi-local fields are symmetric.}.
It can be easily checked that 
no interaction in \eq{eq:fullS}
gives rise to a mixing between
modes with different relative momenta
except for the modes with strictly zero center of mass momentum. 
This is because the vertex  in \eq{eq:cubicV}
and all other vertices in  \eq{eq:fullS} are invariant under 
$k$-dependent $U(1)$ tranformations,
$\delta T_{2}^{k_1,k_2} \rightarrow \delta T_{2}^{k_1,k_2} 
e^{i ( \varphi_{k_1} + \varphi_{k_2})}$,
where $\varphi_k$ is $k$-dependent phase angle 
with $\varphi_{-k} = -\varphi_{k}$.
These $U(1)$ symmetries forbid mixing between
modes with different relative momenta.

 \begin{figure}[ht]
 \begin{center}
 \includegraphics[scale=0.4]{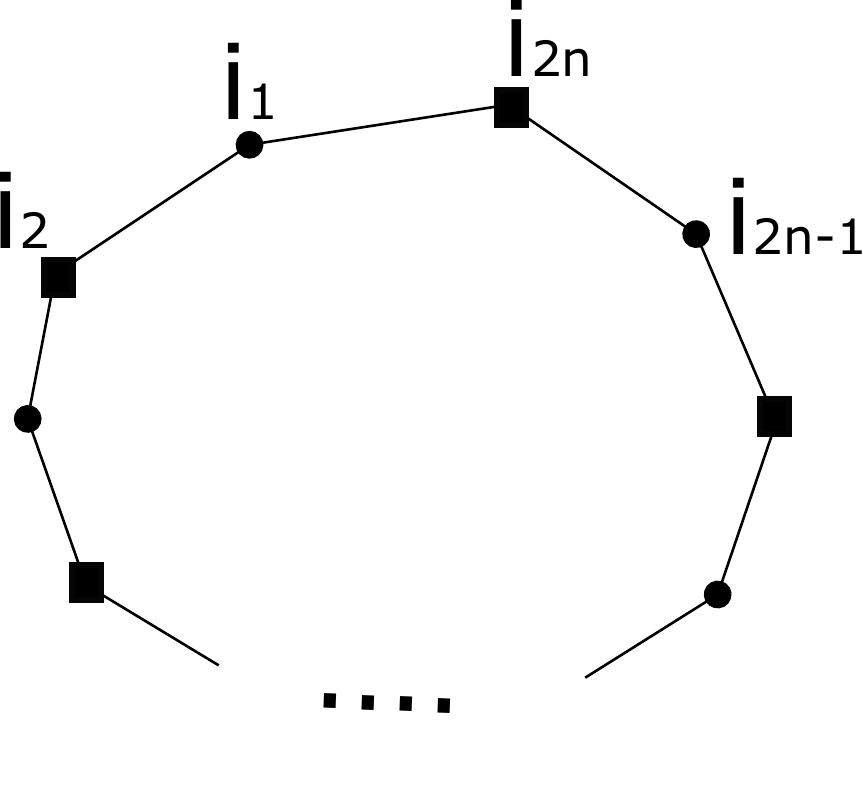} 
  \end{center}
 \caption{
With the lower flavour symmetry group of
$O(N/2) \times O(\sqrt{N/2}) \times O(\sqrt{N/2})$,
additional operators are allowed in the basis states 
that span the kinematic Hilbert space.
Besides the bi-local operators included in \eq{eq:symm}, 
one has to include
$W^C = 
\left( \frac{N}{2} \right)^{-\frac{n-1}{2}}
\tr{ 
\pi^{i_1} 
(\pi^{i_2})^T 
\pi^{i_3} 
(\pi^{i_4})^T 
...
\pi^{i_{2n-1}} 
(\pi^{i_{2n}})^T 
}$
defined on a series of sites
$C=(i_1,i_2,....,i_{2n})$
that can be viewed as a loop,
where $\pi^i$ is is the matrix 
obtained by rearranging $N/2$ components of $\Pi_A^{~i}$ with $A=L+N/2+1,..,L+N$
into  a $\sqrt{N/2}$  by $\sqrt{N/2}$ matrix. 
In the figure, circles (squares) denote sites 
with $\pi$ ($\pi^T$).
 }
 \label{fig:loop_operator}
 \end{figure}

In order to generate mixing between modes with different relative momenta,
one has to break these $U(1)$ symmetries.
One simple way of achieving this is to enlarge the kinematic Hilbert space 
from \eq{eq:symm} to the one  
in which the $O(N/2)  \times O(N/2)$ flavour symmetry is further broken down to 
$O(N/2)  \times O(\sqrt{N/2}) \times O(\sqrt{N/2})$\footnote{
Here, $N=M-L$ is chosen such that $\sqrt{N/2}$ is an integer.}.
The first $O(N/2)$ acts on 
$\Pi_A^{~i}$ with $A=L+1,..,L+N/2$ as before.
The remaining
$O(\sqrt{N/2}) \times O(\sqrt{N/2})$ 
acts on $\Pi_A^{~i}$ with
 $A=L+N/2+1,..,M$
as left and right $O(\sqrt{N/2})$ multiplications 
as $\Pi_A^{~i}$ is viewed as a matrix.
Namely, we identify
$N/2$ components of  $\Pi_A^{~i}$ with $A=L+N/2+1,..,M$
as a $ \sqrt{N/2} \times  \sqrt{N/2}$  matrix :
$\Pi_A^{~i}=\pi_{ {\bf a} {\bf b}}^{~i}$
with 
${\bf a} =\lfloor  \frac{A-(L+N/2+1)}{\sqrt{N/2}} +1 \rfloor$
and
${\bf b} = \left[ A-(L+N/2+1) \mod \sqrt{N/2} \right] + 1$
for $A=L+N/2+1,..,M$.
Under  $O(\sqrt{N/2}) \times O(\sqrt{N/2})$,
$\pi^{~i}$ is transformed as 
$\pi^i \rightarrow o_L \pi^i o_R$,
where $o_L$ and $o_R$ are $\sqrt{N/2} \times \sqrt{N/2}$ 
orthogonal matrices.
%
The enlarged sub-Hilbert space 
with the lower flavour symmetry
is spanned by a larger set of basis states given by
\bqa
\cstrdualc & = & 
\int D \Pi ~ e^{ - i  \left[
\sqrt{N} \sum_{a=1}^L 
  \Pi^{i}_{~a} 
q^a_{~i} 
+ 
\sum_{b=L+1}^{L+\frac{N}{2}} 
 P_{1,ij} 
\Pi_b^{~i} \Pi_b^{~j}
+
 P_{2,ij} 
\tr{ \pi^{~i} (\pi^{~j})^T }
\right]
  }   \times \nn
  &&
 ~~~~~ e^{-i 
  \sum_C X_C W^C
  }
    \cb \Pi \rb.
\label{eq:symm2}
\eqa
The first line of \eq{eq:symm2}
is exactly the same as \eq{eq:symm}
because 
$
\tr{ \pi^{~i} (\pi^{~j})^T }
=
\sum_{b=L+N/2+1}^{M} 
\Pi_b^{~i} \Pi_b^{~j}$.
$q$ is an $L \times L$ matrix
and $P_1$ and $P_2$ are $L \times L$ symmetric matrices as before.
The second line includes additional operators 
that are allowed due to the lowered flavour symmetry.
Besides what is already included in the first line,
the most general operators needed to span the 
Hilbert space with $O(\sqrt{N/2}) \times O(\sqrt{N/2})$ symmetry
are the Wilson-loop-like operators
$W^C = 
\left( \frac{N}{2} \right)^{-\frac{n-1}{2}}
\tr{ 
\pi^{i_1} 
(\pi^{i_2})^T 
\pi^{i_3} 
(\pi^{i_4})^T 
...
\pi^{i_{2n-1}} 
(\pi^{i_{2n}})^T 
}$
defined on a series of sites $C=(i_1,i_2,....,i_{2n})$
(see \fig{fig:loop_operator})\cite{Ma:2022aa},
where the prefactor 
normalizes the multi-site loop operators as 
$W^C \sim O(N)$ in the large $N$ limit.
In the new term, we only include loop operators with $n \geq 2$
because the bi-local operators,
which are the special case of $W^C$ with $n=1$, 
are already included in the first line.
The enlarged kinematic Hilbert space 
is spanned by the bi-local fields
and the new multi-local fields $X_C$ 
which are defined in the space of loops.

 \begin{figure}[ht]
 \begin{center}
     \subfigure[]{
 \includegraphics[scale=0.4]{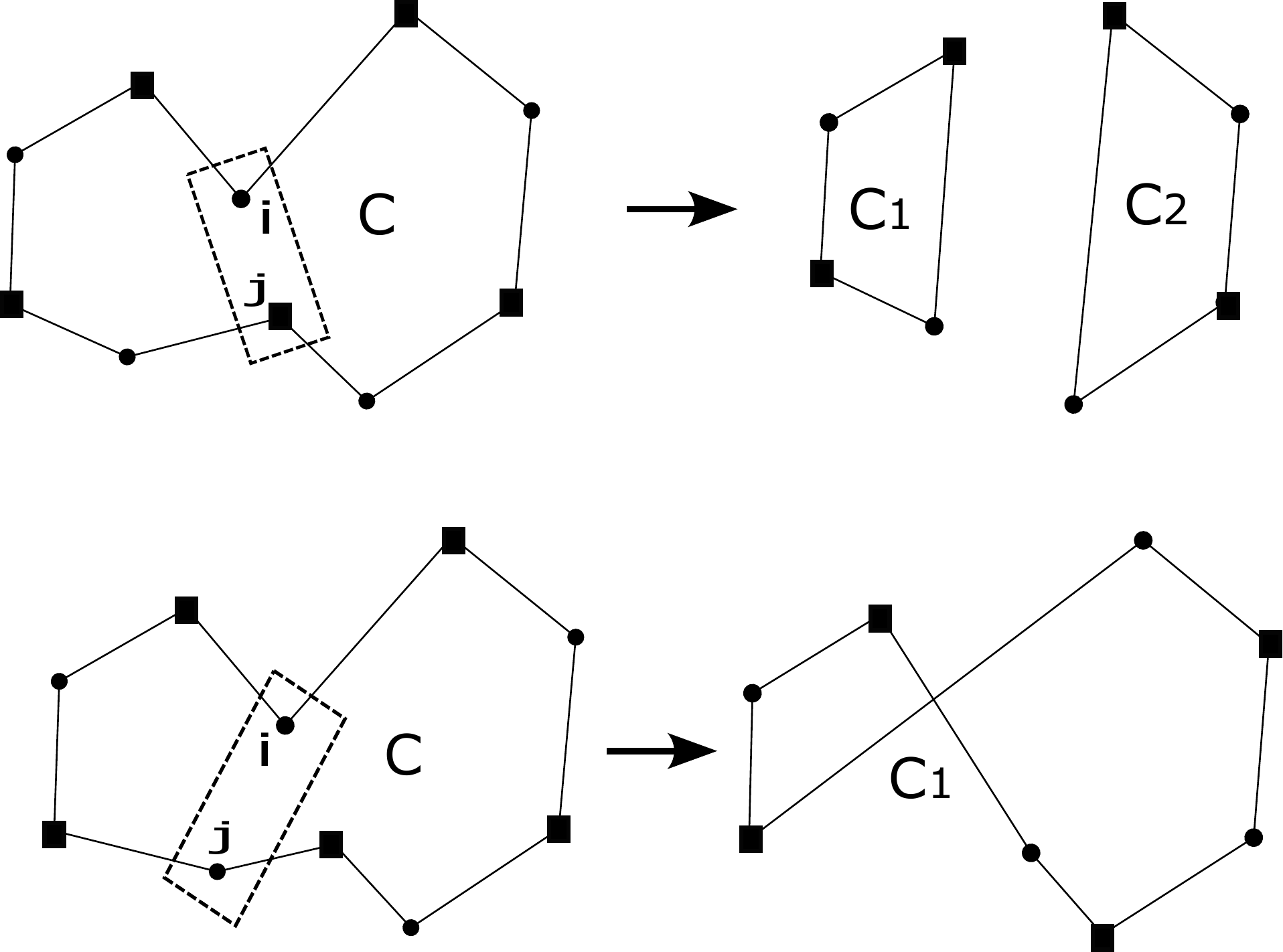} 
  \label{fig:loop1}
 } 
  \hfill
 \subfigure[]{
 \includegraphics[scale=0.45]{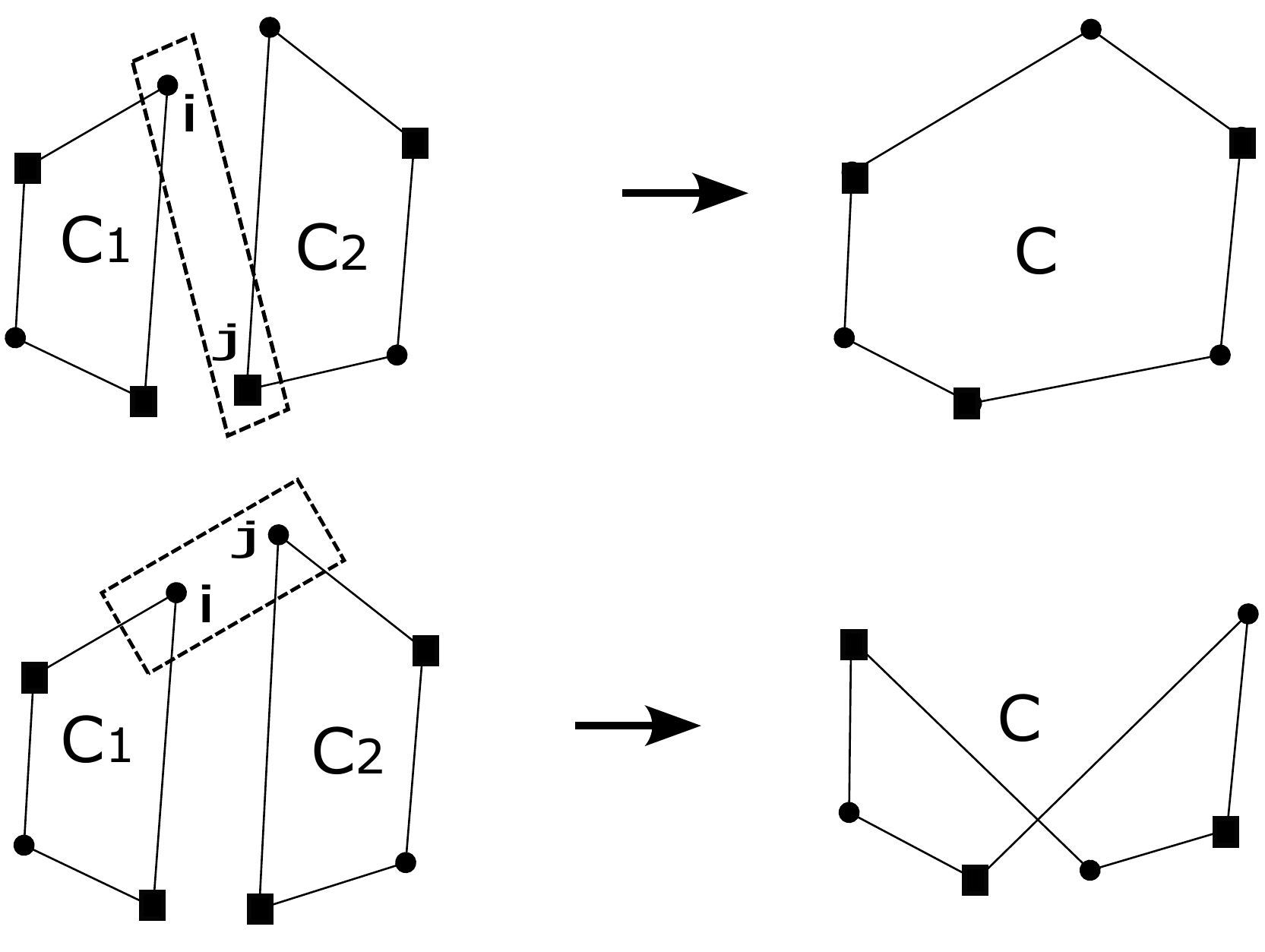} 
 \label{fig:loop2}
 } 
  \end{center}
 \caption{
 The dynamics that $Q_{ij}$ induces on loop operators.
 (a) At the linear order in the source of loop ($X_C$), 
 a loop either splits into two loops or shrinks to a smaller loop. 
 If $\pi$ and $\pi^T$ are contracted, one loop is broken into two loops.
 If two $\pi$'s (or two $\pi^T$'s) are contracted, 
 one of the segment is reversed before glued
 to the other segment to form one smaller loop.
 (b) At the quadratic order in the source, 
 two loops merge into one loop. 
 If $\pi$ and $\pi^T$ are contracted, two loops merge 
 without changing their orientations.
If two $\pi$'s (or two $\pi^T$'s) are contracted, 
 one of the segment is reversed before merging.
  }
 \label{fig:loop12}
 \end{figure}

Because $\{ \cstrdualc \}$ forms a complete basis 
of the Hilbert space with the symmetry,
$e^{ - i \epsilon ( \hat H_{v_0}  + \hat G_{y_0} ) } \cstrdualc$
can be expressed as a linear superposition of $\cstrdualc$.
The theory of the new set of collective variables can be derived
in the same way that  \eq{eq:fullS} is derived,
\bqa
S &=&  N \int d\tau ~ \Bigl[
\tr{  
s \partial_\tau q
+T_1 \partial_\tau P_1
+T_2 \partial_\tau P_2 }
+ \sum_C Y^C \partial_\tau X_C \nn
&&
-\tr{ \cH[q,s,P_1,T_1,P_2,T_2,X,Y] v  
+  \cG[q,s,P_1,T_1,P_2,T_2,X,Y] y  } \Bigr].
\label{eq:fullSX}
\eqa
Here $X_C$ is the dynamical source for the loop operator
just as $P_1$ and $P_2$ are promoted to dynamical variables
in \eq{eq:fullS}.
$Y^C$ is the conjugate momentum of $X_C$
whose saddle-point value represents the expectation value of the loop operators, 
$\lb Y^C \rb = \frac{1}{N} \lb W^C \rb$.
The momentum constraint and Hamiltonian constraints are modified to
\bqa
\cG^j_{~i}[q,s,P_1,T_1,P_2,T_2,X_C,Y^C]  &=&  
 \left[ s q  +  2 ( T_1 P_1 + T_2 P_2 )
+  i \beta  I  \right]^j_{~i}
  + \sum_C Y^{C-i+j} X_C, \nn
 \cH[q,s,P_1,T_1,P_2,T_2]  &=& 
-U + \ta U Q U.
 \label{HG2}
 \eqa
The last term in the momentum constraint is
the new addition that describes the action 
of a generalized diffeomorphism under which
loop $C$ is deformed into a new loop $C-i+j$ 
which is obtained by removing 
site $i$ with $j$ in $C$.
If $C$ does not include $i$, $Y^{C-i+j}=0$.
In the Hamiltonian,
$ U^{ij} = 
  \left( s s^T + T_1 + T_2 \right)^{ij}$ is unchanged,
but $Q_{ij}$ is modified with additional terms
that involve general loop fields,
\bqa
  Q_{ij} & =&   
 \left( q^T q + \sum_c [ 4  P_c T_c P_c + i P_c ] +
 2 i \sum_C X_C \sum_{C_1,C_2} F^C_{C_1,C_2;ij} Y^{C_1} Y^{C_2} \right. \nn
 && \left.
 + 2 \sum_C X_C 
\sum_l
 \left[
 G^C_i P_{2,jl} Y^{C-i+l} 
+ G^C_j P_{2,il} Y^{C-j+l} 
 \right] 
 + \sum_{C_1,C_2} X_{C_1} X_{C_2} G^{C_1,C_2}_{C;ij} Y^C
 \right)_{ij}.
 \label{eq:newQ}
 \eqa
Here, the first two terms are the same as before.
The third term describes the process where
loop $C$ breaks into loops $C_1$ and $C_2$
with the removal of sites $i$ and $j$ out of $C$ and 
rejoining the remaining segments.
The way the remaining segments are rejoined depends on 
whether $i$ and $j$ are separated 
by an even or odd number of sites. 
If $C$ includes both $i$ and $j$,
it can be written as  $C =  i + C' + j + C''$ without loss or generality, 
where $C', C''$ represent open chains 
that form loop $C$ once $C'$ and $C''$ 
are glued via $i$ and $j$.
Let $n_{C'}$ denote the number of sites in chain $C'$.
If $n_{C'}$ is even,
$F^C_{C_1,C_2;ij}=1$
for $C_1=C'$ and $C_2=C''$.
If $n_{C'}$ is odd,
$F^C_{C_1,C_2;ij}=\sqrt{2/N}$
for $C_1=C'+ \bar C''$
and $C_2=\emptyset$.
Here,  $\bar C''$ denotes the chain 
constructed by reversing the order of sites in $C''$.
For the loop made of the empty set,
we use the convention of $Y^\emptyset \equiv 1/2$.
This is illustrated in   \fig{fig:loop1}.
While $n_C \geq 4$,
$n_{C_1}$ and $n_{C_2}$ can be 
any non-negative even integer
because loops generated from $C$ can be of smaller sizes.
For example, $C_2=\emptyset$
if $i$ and $j$ are adjacent in $C$. 
If $C_2$ is bi-local with $C_2=(ij)$,
$Y^{C_2} \equiv T_2^{ij}$.
If $C$ does not include $i$ or $j$, $F^C_{C_1,C_2;ij}=0$.
The fourth term describes the process where
loop $C$ merge with a bi-local field $P_{2}$ to create a new loop
by replacing site $i$ from $C$ and site $j$ from $P_2$,
or vice versa.
If $C$ includes site $i$, $G^C_i=1$.
Otherwise, $G^C_i=0$.
$C-i+l$ represent the loop obtained by replacing 
site $i$ with $l$ in $C$. 
In the last term,
loops $C_1$ and $C_2$ 
merge into a new loop $C$
by removing a site from each loop and rejoining them.
If $C_1$ and $C_2$ include site $i$ and $j$, respectively,
we can write
$C_1 = i + C_1'$
and 
$C_2 = j + C_2'$.
If site $i$ has $\pi$  and $j$ has $\pi^T$ (or vice versa),
$G^{C_1,C_2}_{C;ij}=1$ 
for 
$C=C_1' + C_2' $.
If site $i$ and $j$ both have $\pi$ (or $\pi^T$),
$G^{C_1,C_2}_{C;ij}=1$ 
for 
$C=C_1' +\bar C_2' $.
This is illustrated in   \fig{fig:loop2}.
The induced dynamics of loops is similar 
to the dynamics that loop fields obey 
in holographic duals of lattice gauge theories\cite{Lee2012}.
It is noted that the second to the last term can be
viewed as a special case of the last term
where one of the merged loops is just bi-local.

 \begin{figure}[ht]
 \begin{center}
 \includegraphics[scale=1]{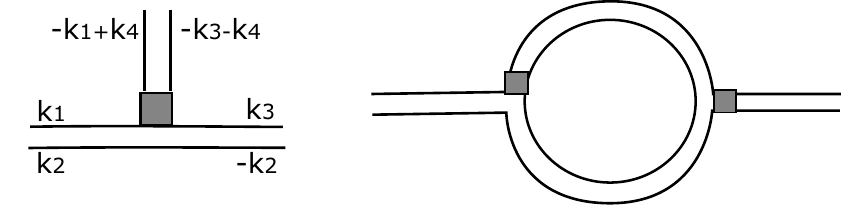} 
  \end{center}
 \caption{
The diagram on the left shows a new cubic vertex for the bi-local field
$\delta T_2$ in the presence of a non-zero expectation value
of the four-site loop field $\bar X_{k1,k2,k3,k4}$ as is shown in  
\eq{eq:cubicV2}.
Unlike the vertex in \label{fig:cubic1},
the momentum in each single line does not have to be conserved
because the four-site loop field breaks the local $Z_2$ symmetry.
Consequently, the one-loop self-energy shown in the right panel 
has a non-zero off-diagonal element between bi-local fields 
with different relative momenta (see \eq{eq:self2}).
 }
 \label{fig:cubic2}
 \end{figure}

The semi-classical equation of motion for $U$ and $\cs$,
which determine the metric,
remains the same as \eq{eq:dtau us} 
even in the presence of the additional loop fields.
This is because $U$ depends only on the bi-local fields
$(ss^T)$, $T_1$ and $T_2$.
Therefore, the equation of motion for the spin-$2$ modes remains
the same and there still exist a continuum of spin-$2$ modes 
labeled by the relative momentum of the bi-local fields in the large $N$ limit.
However,  differences arise from $1/N$ corrections
because the general loop-fields give rise 
to new interaction vertices.
For example, 
$\ta U Q U$ in \eq{HG}
generates a cubic vertex for $\delta T_2$,
$  i \alpha \sum_{ijkl}
 \delta T_2^{ij} \bar X_{jklm}  \delta T_2^{ki} \delta T_2^{lm}$,
 where $\bar X_{i_1 i_2 i_3 i_4}$ represents the saddle-point value
of the four-site loop field.
In momentum space, this gives rise to a vertex 
that breaks the $k$-dependent $U(1)$ symmetry,
\bqa
i \ta  \int dk_1 dk_2 dk_3 dk_4 ~
\bar X_{-k_1,-k_3,k_1-k_4,k_3+k_4}
~
\delta T_{2}^{k_1,k_2} \delta T_{2}^{-k_2,k_3} \delta T_{2}^{-k_1+k_4,-k_3-k_4}.
\label{eq:cubicV2}
\eqa
Without the $k$-dependent $U(1)$ symmetry,
loop-corrections can give rise to 
the self-energy that is off-diagonal in the space of relative momentum.
Through the one-loop correction shown in  \fig{fig:cubic2}, 
one obtains the self-energy for $\delta T_2$,
\bqa
\Sigma_{k_1,k_2; k_1-l, k_2+l } 
\sim
\ta^2 \int dq ~
\bar X_{-k_1,-q,q,k_1}
\bar X_{-k_1, -k_2, k_1-l, k_2+l}
G_2(k_1,q) G_2(q,-k_2).
\label{eq:self2}
\eqa 
While the center of mass momentum is still conserved,
the self-energy mixes modes with different relative momenta.
This off-diagonal self-energy also creates mixing 
between 
$\delta U_{\frac{k+q}{2},\frac{k-q}{2}}$
and 
$\delta U_{\frac{k+q'}{2},\frac{k-q'}{2}}$
for $q \neq q'$
because $\delta T_2$  linearly mixes with $\delta U$. 
In the presence of such mixings, 
the eigenmodes of the wave equation 
derived from the quantum effective action
should be given by
linear superpositions of modes with 
different relative momenta as
\bqa
\delta U^{(l)}_k = \int dq ~ f^{(l)}_{k,q} \delta U_{\frac{k+q}{2},\frac{k-q}{2}},
\eqa
where each eigenmode is labeled by the center of mass momentum $k$
and an additional label $l$.
Finding the eigenvector $ f^{(l)}_{k,q}$ reduces to the problem of 
diagonalizing a quantum mechanical Hamiltonian of a `particle' 
moving in the space of relative momentum. 
The particle is subject to a potential $N m_q^2$
because of the mass term that is diagonal 
in relative momentum (see \eq{eq:gravitonEOMq}).
The off-diagonal self-energy allows the particle 
to hop from $q$ to $q'$ 
with hopping amplitude proportional to
$\Sigma_{ \frac{k+q}{2},\frac{k-q}{2}; \frac{k+q'}{2},\frac{k-q'}{2}}$.
The diagonalization of the Hamiltonian 
will give rise to a discrete set of bound states at low energies 
because the $N m_q^2$ provides a harmonic potential at low $q$.
The true graviton should stay gapless 
due to the diffeomorphism invariance
and the unbroken translational invariance.
However, other spin-$2$ modes are expected to 
acquire non-zero masses that are order of $1/N$ 
as their masses are not protected from quantum corrections.

\section{Discussion}
\label{sec:summary}

In this paper, we show that the model of quantum gravity
proposed in Ref. \cite{Lee:2020aa} supports a gapless spin-$2$ excitation
as a propagating mode.
Although the model has no pre-determined partitioning of the Hilbert space
into local Hilbert spaces, the low-energy effective theory 
takes the form of a local theory with an emergent Lorentz symmetry
in a frame where 
the pattern of entanglement exhibits a local structure
and local clocks are well defined.
We conclude with some open questions.
First, the present model has  
a continuum of spin-$2$ modes 
with a continuously varying mass in the large $N$ limit.
This unrealistic feature is expected to go away
once the kinematic Hilbert space is enlarged
and $1/N$ corrections are included as is discussed 
in the previous section. 
It will be of interest to 
take into account all leading $1/N$ corrections
and compute the full 
mass spectrum of the propagating modes.
However, this wouldn't be fully satisfactory 
in that there are still light massive spin-$2$ modes 
in the semi-classical limit.
It is desirable to find a new mechanism 
that isolates the massless graviton 
from other massive modes with a mass gap 
that is not suppressed in the large $N$ limit.
Second, the present theory suffers from 
the cosmological constant problem.
Without fine tuning, there is no separation between
the scale that controls the rate at which 
time dependent background fields change
and the scale that suppresses higher derivative terms 
in the effective theory. 
It would be interesting to consider an alternative model
(possibly a supersymmetric model)
that stabilizes the flat spacetime as a saddle point.
Despite these drawbacks, this model serves as
a concrete toy model of quantum gravity
that realizes some interesting features 
that the true theory of quantum gravity may share.
Those features are the Hilbert-space-partition-independence
and the emergence of dimension, topology, signature 
and geometry of spacetime.
Finally, we comment on the relation between 
the present  model and the BFSS/IKKT matrix models
that have been proposed 
as  a non-perturbative formulation of string theory\cite{PhysRevD.55.5112,ISHIBASHI1997467,physics5010001}.
Those matrix models share the same goal of realizing
emergent spacetime from non-geometric microscopic degrees of freedom.
However, one notable difference is the fact that  
the number of non-compact spacetime directions is bounded by the number of matrices
in the previous matrix models.
In the present model, the spacetime dimension is dynamical,
and there are states that exhibit spacetimes with any dimension.
It would be interesting to know if there is any relation between
the earlier matrix models and
 the present model restricted to a sub-Hilbert space with a fixed spacetime dimension.
 Ultimately, it will be great to understand a dynamical mechanism
that selects certain spacetime dimensions in the model where 
the spacetime dimension is fully dynamical.

\section*{Acknowledgments}

The research was supported by
the Natural Sciences and Engineering Research Council of
Canada.
Research at Perimeter Institute is supported in part by the Government of Canada through the Department of Innovation, Science and Economic Development Canada and by the Province of Ontario through the Ministry of Colleges and Universities.

 \bibliography{references}
 
\end{document}